\documentclass{article}

\usepackage[english]{babel}
\usepackage[sorting=none]{biblatex}
\usepackage{csquotes}
\usepackage[letterpaper,top=2cm,bottom=2cm,left=3cm,right=3cm,marginparwidth=1.75cm]{geometry}

\usepackage{amsmath}
\usepackage{graphicx}
\usepackage[colorlinks=true, allcolors=blue]{hyperref}
\usepackage[nolist]{acronym}
\usepackage{fontawesome5}
\usepackage[dvipsnames]{xcolor}
\usepackage{float}

\usepackage{ragged2e} 
\usepackage{caption}

\usepackage{array}
\usepackage{booktabs}
\usepackage{tabulary}

\usepackage{microtype} 

\usepackage{tikz}
\usetikzlibrary{arrows,arrows.meta,calc,decorations,decorations.pathmorphing,decorations.pathreplacing,fit,positioning,shadows,shapes,shapes.multipart}
\tikzset{
    node distance=2.75em,
    box/.style={draw, minimum height=1.75em, fill=white},
    partition/.style={box, minimum width=5.5em,
        append after command={
                [shorten <= -0.5\pgflinewidth]
                ([shift={(-0.5\pgflinewidth,-0.5\pgflinewidth)}]\tikzlastnode.north east)
            edge([shift={(-0.5\pgflinewidth,-0.5\pgflinewidth)}]$(\tikzlastnode.north east)+(0,2.5em)$)
                ([shift={(+0.5\pgflinewidth,-0.5\pgflinewidth)}]\tikzlastnode.north west)
            edge([shift={(+0.5\pgflinewidth,+0.5\pgflinewidth)}]$(\tikzlastnode.north west)+(0,2.5em)$)
            }
    },
    interpreter/.style={partition, minimum width=5em},
    tool/.style={box, ellipse},
    artifact/.style={box, rectangle},
    exec env/.style={box, chamfered rectangle, minimum width=6em},
    runtime/.style={box, rounded corners, inner xsep=0.75em, inner ysep=1em, yshift=0.125em},
    runtime title/.style={box, rounded rectangle},
    must know/.style={-stealth, dashed, thin},
    trust/.style={rectangle split, rectangle split parts=2},
    part/.style={draw, minimum width = 8mm, minimum height = 10mm, inner sep = 1.5mm},
    part-container/.style={node distance = 2mm},
    part-label/.style={fill = none, draw = none, inner sep = 0},
    port/.style={draw, rounded corners, thick, -stealth},
    flow/.style={draw, rounded corners, -stealth},
    reconf-flow/.style={very thick, double distance = 1pt, arrows = {-Latex[length=0pt 3 0]}, shorten <= 5mm, shorten >= 5mm},
}

\pgfdeclarelayer{bg}    
\pgfsetlayers{bg,main}  

\usepackage[dvipsnames]{xcolor}
\usepackage{listings}

\lstdefinestyle{C}{language=C, morekeywords={__asm__, _Bool, _Atomic, __attribute__, int64_t, uintptr_t, uint8_t, size_t}}

\usepackage{fontspec}
\usepackage{inconsolata}
\makeatletter
\newcommand{\requirement}[2]{
    \noindent%
    \protected@edef\@currentlabelname{REQ-#1}%
    \phantomsection%
    \label{req:#1}%
    \textbf{REQ-#1}: #2\\%
}
\makeatother

\newcommand{\begincolorgroupA}{\begingroup\color{black}} \newcommand{\colA}{\aftergroup\begincolorgroupA}
\newcommand{\begincolorgroupB}{\begingroup\color{red}} \newcommand{\colB}{\aftergroup\begincolorgroupB}
\newcommand{\begincolorgroupC}{\begingroup\color{Orange}} \newcommand{\colC}{\aftergroup\begincolorgroupC}
\newcommand{\begincolorgroupD}{\begingroup\color{ForestGreen}} \newcommand{\colD}{\aftergroup\begincolorgroupD}
\newcommand{\begincolorgroupE}{\begingroup\color{blue}} \newcommand{\colE}{\aftergroup\begincolorgroupE}
\newcommand{\begincolorgroupF}{\begingroup\color{Plum}} \newcommand{\colF}{\aftergroup\begincolorgroupF}
\newcommand{\begincolorgroupG}{\begingroup\color{RubineRed}} \newcommand{\colG}{\aftergroup\begincolorgroupG}
\def\endcol{\aftergroup\endgroup} 

\title{Beyond Source Compatibility: \\ WebAssembly as a Stable ABI for ARINC~653 Applications}

\usepackage{orcidlink}
\usepackage[affil-it]{authblk}
\newcommand{\authorr}[4]{\author[#2]{
    \href{mailto:#3}{#1}%
    \if\relax\detokenize{#4}\relax%
      \else%
        \textnormal{\orcidlink{#4}}%
      \fi%
    }
}

\authorr{Wanja Zaeske}{1}{wanja.zaeske@dlr.de}{0000-0002-1427-2627}
\authorr{Patrick Siegl}{2}{patrick.siegl@airbus.com}{0009-0007-1630-1595}
\authorr{Florian Hartung}{1}{florian.hartung@dlr.de}{0009-0006-6213-7231}
\authorr{Fabian Bauriedl}{3}{fabian.bauriedl@airbus.com}{}
\authorr{Nicolaus Bär}{3}{nicolaus.baer@airbus.com}{}
\authorr{Umut Durak}{1}{umut.durak@dlr.de}{0000-0002-2928-1710}

\affil[1]{German Aerospace Center (DLR), Institute of Flight Systems, Lilienthalplatz 7, 38108 Braunschweig, Germany}
\affil[2]{Airbus Defence and Space GmbH, Claude-Dornier-Straße, 88090 Immenstaad am Bodensee, Germany}
\affil[3]{Airbus Defence and Space GmbH, Rechliner Straße, 85077 Manching, Germany}

\begin{acronym}[TDMA]
    \acroindefinite{MMU}{an}{a}
    \acro{ABI}{Application Binary Interface}
    \acro{AMC}{Acceptable Means of Compliance}
    \acro{AOT}{Ahead-of-time}
    \acro{APEX}{APplication EXecutive}
    \acro{API}{Application Programming Interface}
    \acro{BaaS}{Backend-as-a-Service}
    \acro{CAN}{Controller Area Network}
    \acro{CAST}{Certification Authorities Software Team}
    \acro{CI}{Continuous Integration}
    \acro{CLI}{Command Line Interface}
    \acro{CPU}{Central Processing Unit}
    \acro{DAL}{Design Assurance Level}
    \acro{DMA}{Direct Memory Access}
    \acro{DevOps}{Development Operations}
    \acro{EASA}{European Union Aviation Safety Agency}
    \acro{FFI}{Foreign Function Interface}
    \acro{FaaS}{Function-as-a-Service}
    \acro{I/O}{Input/Output}
    \acro{IDAL}{Item Development Assurance Level}
    \acro{IDL}{Interface Definition Language}
    \acro{IMA}{Integrated Modular Avionics}
    \acro{IPC}{interprocess communication}
    \acro{ISA}{Instruction Set Architecture}
    \acro{JIT}{Just-in-time}
    \acro{MCP}{Multi-Core Processor}
    \acro{MCS}{Mixed Criticality System}
    \acro{MMU}{Memory Management Unit}
    \acro{MVP}{Minimum Viable Product}
    \acro{OS}{Operating System}
    \acro{PAFA}{Plug and Fly Avionics}
    \acro{PIC}{Position-Independent Code}
    \acro{PID}{Proportional–integral–derivative controller}
    \acro{POSIX}{Portable Operating System Interface}
    \acro{RDC}{Remote Data Concentrator}
    \acro{RTOS}{Real-time operating system}
    \acro{SCP}{Single-Core Processor}
    \acro{SIMD}{Single Instruction, Multiple Data}
    \acro{SRAM}{Static Random-Access Memory}
    \acro{SWAP-C}{Size, Weight, Power and Cost}
    \acro{SoC}{System on a Chip}
    \acro{TLS}{Thread-Local Storage}
    \acro{VM}{Virtual Machine}
    \acro{WAMR}{WebAssembly Micro Runtime}
    \acro{WASI}{WebAssembly System Interface}
    \acro{WAT}{WebAssembly Text Format}
    \acro{WCET}{worst-case execution time}
    \acro{WIT}{WebAssembly Interface Type}
    \acro{Wasm}{WebAssembly}
\end{acronym}

\makeatletter
\newcommand{\implchoice}[2]{
    #1. #2.
}
\makeatother

\begin{document}
\emergencystretch 3em 
\maketitle

\begin{abstract}
Thirty years ago, ARINC~653 was introduced to provide a vendor-independent platform for software applications in \ac{IMA}.
A platform is characterized by a well-defined interface abstracting over implementation details; in the case of ARINC~653 this is an \ac{API}, defined on the source-code level.
However, several aspects of this \ac{API} --- such as the memory layout and partially also the integer bit width of data types --- remain implementation dependent, so portability across \acp{OS} requires recompilation, and often also source code adaptation. 
Our contribution is three-fold;
first, we refine the ARINC~653 header file, eliminating remaining ambiguities.
Second, building on \ac{Wasm}, we define an \ac{OS}- and processor-architecture-agnostic \ac{ABI} at the object-code level, resolving differences in data layout, endianness, and function-pointer representation between host and guest environments.
Third, we demonstrate a \ac{Wasm}-\acs{VM}-agnostic shim-library that executes ARINC~653 partitions, compiled to \ac{Wasm} bytecode and following our \ac{ABI}, on \textit{multiple} processor architectures using two independent \ac{Wasm} runtimes.
\end{abstract}
\acresetall

\section{Introduction}
\label{sec:introduction}
%
%

With the push towards \ac{IMA}, multiple software applications are executed together on the same computer module.
Between applications, partitioning is of utmost importance to avoid cascading failures~\cite{do178c}.
This partitioning is provided by specialized \acp{OS}, offering an \ac{API} for scheduling, \acl{IPC}, external I/O, health management, etc.
Software applications are often tightly coupled to an \ac{OS} and hardware / module type, hindering portability between such \acp{OS}.
To enable software re-use, ARINC~653 was developed~\cite{COOK1997479}.

ARINC~653 is a modular standard that specifies \ac{OS} interfaces and configuration aspects for the partitioning.
Multiple \acp{OS} implement (subsets of) the ARINC~653 \ac{API}, e.g. PikeOS, VxWorks, DEOS, LynxOS-178 \& Integrity-178.
However, various aspects of the ARINC~653 \ac{API} are implementation dependent; for example, the order of struct fields or the bit width of \textit{id} types (such as \lstinline|APEX_LONG_INTEGER|).
Consequently, even applications relying only on the specified behavior of the ARINC~653 \ac{API} still require recompilation for a port to a new compliant \ac{OS} to accommodate these implementation-dependent aspects~\cite{COOK1995575}.
Although Cook~\cite{COOK1995575} frames this as \textit{``[...] facilitating portability via recompilation of the application source [...],''} recompilation is necessary but insufficient to guarantee portability: Cook presupposes access to source code, a matching toolchain, and the absence of implicit reliance on non-standardized behavior.
In practice, applications are likely to implicitly rely on such non-standardized aspects, further raising the risk and cost of porting to a new \ac{OS}.
In particular in \ac{IMA} settings, depending on history and relationship between function supplier and airframer, the source code might not be available (any more) to the airframer.

In short, ARINC~653 enables (but does not guarantee!) source-code level compatibility without making any promises on the object code level except for symbol names.
To address this, we define a new execution environment equipped with an ARINC~653 \ac{API}, based on \ac{Wasm}.
\ac{Wasm} provides a processor-\acs{ISA}-independent, highly standardized, executable bytecode format.
Existing ARINC~653 C header files are refined, eliminating implementation-dependent aspects in a forward-compatible manner.
This is not a novel idea, and has been briefly discussed already in section 4.b of \cite{10311207}.
However, various minute implementation details are merely glanced over by that publication, leaving it up to this paper to fill in the gaps.

For the demonstration in this paper, we focus on the \ac{Wasm} \acp{VM} Wasmtime~\cite{wasmtime} and \ac{WAMR}~\cite{wamr_github}.
To facilitate reproduction of our results, AIRBUS' Linux-based ARINC~653 shell a653lib\footnote{available at \url{https://github.com/airbus/a653lib}} is used for the demonstration.

Looking beyond individual \ac{OS} ports, next-generation \ac{IMA} envisions virtualization to achieve software-defined avionics, building on earlier work on reconfigurable \ac{IMA} platform architectures~\cite{bieber2009preliminary}.
In this vision, applications can be freely deployed and reconfigured across heterogeneous hardware~\cite{SoftwareDefinedAvionicsFVL}.
A stable, hardware-independent \ac{ABI} --- as contributed in this paper --- is a major enabler to that.

\section{Background}

\subsection{Terminology: Host vs Guest Environment}
\label{sec:host_vs_guest}

Throughout the following text, we refer to the \textit{host environment} as all code compiled to a processor's native \acf{ISA} such as x86\_64.
This includes a \ac{VM} to execute bytecode, glue code as well as the process entry point that the \ac{OS} invokes.

\begin{figure}
    \centering






\begin{tikzpicture}[trim left=(os), trim right=(os)] 
    \begin{scope}[node distance=0.75em]
        
        \node[artifact, align=center] (wasm) {Bytecode};
        
        \node[partition, artifact, below = of wasm, text width=11em, align=center] (vm) {\acf{VM}};

        \node[below = of vm] (hidden) {};

        \node[partition, below = of vm, fit=(vm.west |- hidden) (vm.east |- hidden), text width=16em, label=center:Process] (process) {};
        
        \node[exec env, inner xsep = 0mm, below = of process, text width=16em, align=center, minimum width=8.5em] (os) {\acf{OS}};

        \coordinate (labelheight) at ($(process.east)+(0,0)$);
        \coordinate (guestenvbot) at (labelheight  |- wasm.south);
        \coordinate (guestenvtop) at (labelheight  |- wasm.north);
        \coordinate (hostenvbot) at (labelheight  |- process.south);
        \coordinate (hostenvtop) at (labelheight  |- vm.north);
        \coordinate (osbot) at (labelheight  |- os.south);
        \coordinate (ostop) at (labelheight  |- os.north);
        
        \draw[decorate,decoration={brace,amplitude=5pt,mirror,raise=1em}](guestenvbot) -- (guestenvtop)
        node[pos=0.5,right=1.5em,black]{Guest environment};
        \draw[decorate,decoration={brace,amplitude=5pt,mirror,raise=1em}] (hostenvbot) -- (hostenvtop)
        node[pos=0.5,right=1.5em,black]{Host environment};
        \draw[decorate,decoration={brace,amplitude=5pt,mirror,raise=1em}] (osbot) -- (ostop)
        node[pos=0.5,right=1.5em,black]{OS};
    \end{scope}
\end{tikzpicture}
    \caption{Integration of a \ac{VM} into a generic execution environment provided by an \ac{OS} in the form of a process}
    \label{fig:tech_stack}
\end{figure}
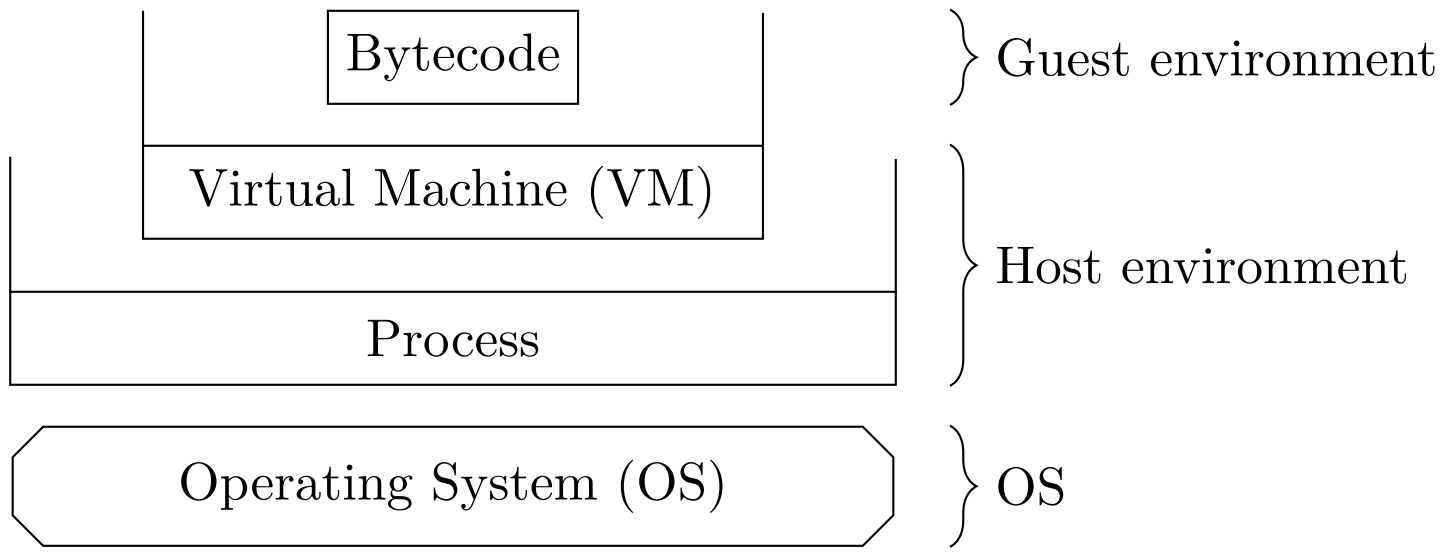

In contrast, the \textit{guest environment} consists of all code compiled into the bytecode, that is executed in the \ac{VM}.
When starting the \ac{VM}, the host environment invokes a designated entry function in the guest environment, thus transferring control flow to the latter.
The host environment may expose functions to the guest environment, allowing the guest environment to call back into the host environment, returning control flow back.
\autoref{fig:tech_stack} illustrates the semantics of host-, guest environment and \ac{OS}.
For further illustration consult \autoref{fig:demo_architecture}, which maps the aforementioned terms to the demonstrator technology stack.
For this paper, we consider the \ac{OS} itself distinct from the host environment, due to the distinct code provenance: an application provider/partition developer can change host- and guest environment, but not \ac{OS} code.

\subsection{ARINC~653}
\label{sec:background:arinc_653}

Development of the ARINC~653 standard started in 1990\cite{COOK1995575}, with the goal of providing an Application Programming Interface (\ac{API}) to complement the emerging trend towards \ac{IMA}.
The base standard ARINC~653 Part 1 specifies interfaces to manage partitions, processes, time, memory, inter- and intra-partition communication as well as health monitoring~\cite{ARINC653P0_2}, without dictating a specific implementation of these.
In that sense ARINC~653 is comparable to \ac{POSIX} --- an \ac{OS} independent \ac{API} to enable development of portable applications.
The original standardization process was governed by six defining features, leading to the release of ARINC~653 Part 1 in 1997~\cite{COOK1995575}:

\begin{enumerate}
    \item \textbf{System integrity}.
        Foremost, multiple applications executing side-by-side on the same \ac{OS} must not interfere with each other, faults must be contained.
        Spatial- and temporal-partitioning, realized via static, mutually exclusive allocation, inhibit unintended fault propagation.
        In addition, \ac{OS}-governed access to shared hardware resources completes the partitioning.
    \item \textbf{Language independence}.
        ARINC~653 itself shall remain independent of a single programming language.
        While the standard describes interfaces in C and Ada, other languages are considered.
        The standardization committee recently added a C++ variant of the \ac{API}.
        External authors have published interfaces in Rust~\cite{Friedrich2022PlatformBasedAvionics}.
    \item \textbf{Hardware independence}.
        Application portability is a concern of ARINC~653, therefore the interfaces must enable application implementation that is independent of the underlying hardware architecture.
    \item \textbf{System communication}.
        Communication between a partition, other partitions and peripheral equipment in the aircraft, is message based, and transparent over the physical medium used.
        This includes physical media like ARINC 429, 629, 659 or 664, as well as communication between partitions resident on the same processor module (i.e. via shared memory managed by the \ac{OS}).
        Drivers and addressing become a responsibility of the \ac{OS} and its configuration; therefore out-of-scope for the partitions themselves.\footnote{In practice however, common ARINC~653 \acp{OS} leak implementation details of hardware related I/O ports, necessitating change in a partition software e.g. to change an ARINC~664/AFDX to an ARINC~429 transport layer.}
    \item \textbf{Health monitoring}.
        Interfaces to monitor each partition's health, and to facilitate appropriate mitigation actions upon fault detection.
    \item \textbf{Support for systems of all criticalities}.
        ARINC~653 is designed to support applications up to Software Level A, the most rigorous recognized by ED-12C/DO-178C~\cite{do178c}.
\end{enumerate}

The standard is divided among multiple Parts (0 to 5), most of which are available as multiple supplements (revisions).
Part 0 contains an overview, the required \ac{API} is described in Part 1 and the optional extended \ac{API} in Part 2, the latter released in 2007.
Compliance tests are specified in Part 3A (for the required \ac{API}) and Part 3B (for the extended \ac{API}).
Part 4 contains a subset of the required services, most notably constrained to a maximum of two processes per partition with radically simplified intra-partition scheduling and removes all intra-partition communication services\cite{ARINC653P4}.
Finally, Part 5 elaborates on the recommended capabilities for the implementation of a compliant \ac{OS}.

The \ac{API} description and system design guidelines are accompanied by code examples.
Function signatures and data types are provided in C and Ada, combined with their nominal behavior in the form of pseudo-code~\cite{ARINC653P0_2}.
Notably, the standard requires partitions to provide a single entry point\footnote{Once initializing, a partition may annunciate additional entry points for the invocation of processes to the \ac{OS}.}~\cite[ch. 2.3.1.4.2.1]{ARINC653P1_5}.
For both C and Ada the partition entry point is defined by the corresponding programming language's standard (i.e. a \lstinline|int main(){}| function for C).

The execution model of ARINC~653 recognizes two layers of scheduling, corresponding to the partition and the process level.
On the inter-partition level, static scheduling based on design-time configurable allocations of time (partition windows) is used.
Nevertheless, each partition can comprise multiple processes, which are scheduled dynamically, priority-preemptive and deadline aware within the partition window.
While partitions are spatially isolated from each other, processes within a partition share the same address space, i.e. feature shared memory with each other~\cite[ch. 2.3.2]{ARINC653P1_5}.
This is in contrast to modern general purpose \acp{OS}, in which processes form execution contexts spatially isolated from one another while multiple threads of a given process enable concurrency with shared memory (see \autoref{tab:arinc653_partition_vs_process} for a comparison)~\cite{ARINC653P1_5}.

\begin{table}[htbp]
  \centering
  \captionsetup{justification=centering}
  \caption{Difference between processes in ARINC~653 vs. Linux}
  \label{tab:arinc653_partition_vs_process}

  \small
  \setlength{\tabcolsep}{6pt}
  \renewcommand{\arraystretch}{1.25}

  \begin{tabulary}{\linewidth}{@{}
    >{\centering\arraybackslash}p{0.20\linewidth}
    >{\centering\arraybackslash}p{0.20\linewidth}
    >{\raggedright\arraybackslash}p{0.5\linewidth}
  @{}}
    \toprule
    \multicolumn{2}{@{}c@{}}{\textbf{Term}} & \multicolumn{1}{>{\centering\arraybackslash}p{0.5\linewidth}}{\textbf{Semantics}} \\
    \cmidrule(lr){1-2}
    \textbf{ARINC~653} & \textbf{Linux} & \\
    \midrule
    Partition & Process & Multiple in one \ac{OS}, each with separate memory \\
    Process   & Thread  & Multiple in one ARINC~653 partition / Linux process, with shared memory \\
    \bottomrule
  \end{tabulary}
\end{table}

\subsection{WebAssembly}
\label{sec:background_wasm}

Originally, \acf{Wasm} was designed as a bytecode to speed up computation on the web by complementing JavaScript~\cite{webassembly}.
However, in contrast to other bytecodes as used for Java or Python, \ac{Wasm} intends to be a compilation target for any programming language, enabling among others C, C++, Rust and Ada to run within the browser~\cite{WebAssemblyCoreSpecification1}.
As such, \ac{Wasm} bytecodes compiled from different languages can be linked against each other.
Today, \ac{Wasm} is widely adopted by major browser vendors~\cite{spidermonkey_website,v8_javascript_engine_website}.
Prior work has motivated its use in other non-web domains, such as edge computing~\cite{wasm_for_edge_computing}, embedded systems~\cite{wasm_potential_embedded_systems}, and avionics~\cite{10311207, zaeske_wasm_avionics_2025}.
The motivation for use in avionics is primarily based on \ac{Wasm}'s strong guarantees regarding portability and isolation~\cite{zaeske_wasm_avionics_2025}.


\ac{Wasm} is defined as a virtual, stack-based \ac{ISA}~\cite{webassembly, WebAssemblyCoreSpecification1}, i.e. an \ac{ISA} that is not executed directly on hardware but instead in a \ac{VM} --- also referred to as a runtime or interpreter.
Furthermore, most instructions, including function calls, use a \textit{value stack} to pass and return values.
The value stack does not allow arbitrary access to its contents via indexing; instead only the top-most value may be accessed at all times.
Furthermore, \ac{Wasm} employs a Harvard architecture, enforcing a strict separation of code and data~\cite{webassembly, WebAssemblyCoreSpecification1}.
Combined, the inaccessibility of the call stack and the strict code/data separation prevent attacks such as return-oriented programming~\cite{Lehmann2020, zaeske_wasm_avionics_2025}.

\subsubsection{Linear Memory}
\label{subsubsec:linear_memory}

A \ac{Wasm} module may define or import a \textit{linear memory}~\cite{WebAssemblyCoreSpecification1}.
A linear memory is a contiguous, grow-able block of bytes over which the \ac{Wasm} code has full control.
\ac{Wasm} code can load and store data by indexing into the memory through 32-bit integers, which are always bounds-checked by the \ac{VM}, preventing out-of-bounds accesses~\cite{webassembly, WebAssemblyCoreSpecification1}.
When compiling languages such as C/C++ or Rust to \ac{Wasm}, pointers and references are lowered to such indices.


\subsubsection{Reference Types}
\label{subsubsec:reference_types}


\ac{Wasm} uses a primitive type system: \textit{values} cover both \textit{numbers} (\texttt{i32}, \texttt{i64}, \texttt{f32}, \texttt{f64}, \texttt{v128}) and \textit{references}, among which are \textit{function references} (\texttt{funcref})~\cite{WebAssemblyCoreSpecification1}.
Notably, a function reference in \ac{Wasm} is not a linear memory address but an integer identifying a specific function.
While numeric values can be stored in linear memory, references cannot.

Instead, reference types are exclusively stored in \textit{tables}~\cite{WebAssemblyCoreSpecification1}.
A table is an indexed array of references, enabling C-style function pointer arrays and, for instance, vtables as generated by C++ compilers.
Function references stored in tables can also be called dynamically via the \lstinline|CALL_INDIRECT| instruction, which indexes into a table and performs a type-checked function call.

In addition to existing reference types, \ac{Wasm} 3.0 introduces various new garbage-collected reference types via the GC proposal~\cite{WebAssemblyCoreSpecification3}.
However, its non-deterministic nature renders it unsuitable for avionics contexts; thus, this paper does not address it further.
For safety critical systems, we recommend a subset of the \ac{Wasm} specifications after version 2 to avoid certification issues.

\subsubsection{\texorpdfstring{\ac{Wasm}}{Wasm} as a Compilation Target}

\begin{table}
\centering
\captionsetup{justification=Centering}
\caption{Multiply by three and increment by one expressed in C, the \acf{WAT} and the \ac{Wasm} binary format}
\label{tab:wasm_formats}
\begin{tabular}{ccc}
\hline
\textbf{C} & \textbf{\ac{WAT}} & \textbf{\ac{Wasm}} \\
\hline
\begin{minipage}[t]{0.25\linewidth}
\begin{lstlisting}[style=c,frame=none,escapechar=\|, numberstyle=\tiny\color{black}, basicstyle=\color{gray}\ttfamily\small]
|\colA|int|\endcol| |\colG|foo|\endcol||\colA|(int x) {|\endcol|
  |\colA|return|\endcol| |\colC|3|\endcol| |\colD|*|\endcol| |\colB|x|\endcol| |\colF|+|\endcol| |\colE|1|\endcol|;
|\colA|}|\endcol|
\end{lstlisting}
\end{minipage}
&
\begin{minipage}[t]{0.38\linewidth}
\begin{lstlisting}[frame=none, escapechar=\|, basicstyle=\color{gray}\ttfamily\small, numberstyle=\tiny\color{black}]
(module 
  |\colA|(func|\endcol| |\colG|(export "foo")|\endcol|
    |\colA|(param i32) (result i32)|\endcol|
    |\colB|local.get 0|\endcol|
    |\colC|i32.const 3|\endcol|
    |\colD|i32.mul|\endcol|
    |\colE|i32.const 1|\endcol|
    |\colF|i32.add|\endcol||\colA|)|\endcol|)
\end{lstlisting}
\end{minipage}
&
\begin{minipage}[t]{0.25\linewidth}
\begin{lstlisting}[frame=none, escapechar=\|, basicstyle=\color{gray}\ttfamily\small, numberstyle=\tiny\color{black}]
0061 736d 0100 0000
0106 01|\colA|60 017f 017f|\endcol|
0302 01|\colA|00|\endcol| 0707 01|\colG|03|\endcol|
|\colG|666f 6f00 00|\endcol|0a 0c01
0a00 |\colB|2000|\endcol| |\colC|4103|\endcol| |\colD|6c|\endcol||\colE|41|\endcol|
|\colE|01|\endcol||\colF|6a|\endcol| |\colA|0b|\endcol|

\end{lstlisting}
\end{minipage}
\\
\hline
\end{tabular}
\end{table} 

The smallest \ac{Wasm} code unit is a \ac{Wasm} \textit{module}.
A module contains various \ac{Wasm} objects such as type definitions, functions, globals, memories, imports, or exports\cite{WebAssemblyCoreSpecification2}.
The types of objects that can be imported and exported are functions, globals, linear memories, and tables.
Imported objects can be provided by exports of other \ac{Wasm} modules or by the host environment; however, at the core \ac{Wasm} level, the linking and resolution process is not specified by the standard and depends on the \ac{VM} used.
In practice, modules exist primarily in their efficient bytecode format or their human-readable \acf{WAT} format (\autoref{tab:wasm_formats}).
Both formats are functionally equivalent and can be converted into each other.

Most of the time, a \ac{Wasm} module is created through compilation from a higher-level language such as C, C++, Rust or Ada, though other means of creation exist.
Given this diversity of source languages and deployment environments, compilation backends for \ac{Wasm} have multiplied beyond the browser and co-evolved with the extensions to the original specification~\cite{zhang_research_2024,serrano_snapshot_2025}.
The combinations of all backend extensions bring about the issue of incompatibility between \ac{Wasm} modules and given \ac{VM}s.
Nonetheless, three \ac{Wasm} targets stand out~\cite{zhang_research_2024,serrano_snapshot_2025}: a compilation target that assumes a browser \ac{API} (wasm-emscripten), one replacing this \ac{API} with general-purpose \acp{OS} utilities for out-of-browser applications (wasm-wasi), and one with no prerequisites from the \ac{VM} environment at all (wasm-unknown).
The LLVM toolchain~\cite{llvm_website} emerges as the most prominent backend option.
It is shared by many language frontends such as clang for C/C++ and rustc for Rust and supports all of the \ac{Wasm} targets via the aid of compiler flags.
Of particular interest for avionics is Ada via its GNAT LLVM backend~\cite{adacore_wasm_support}.

\subsubsection{The Execution Lifecycle}

Before a \ac{Wasm} module binary can be executed, it undergoes several preparatory phases.
The module binary is first \textbf{decoded}: its sections are parsed into a \ac{Wasm} module.
This module is then \textbf{validated}: the \ac{VM} checks type safety of value stack accesses, control flow integrity, and other invariants to guarantee sound and efficient execution~\cite{WebAssemblyCoreSpecification2}.

For \ac{AOT} or \ac{JIT} capable \ac{VM}s, \textbf{compilation} follows, translating the validated module to native machine code.
Notably, a compiled or validated module is reusable and can be \textbf{linked} and \textbf{instantiated} multiple times independently.
During linking, imports and exports across modules are resolved; inter-module dependencies must be acyclic.
Unlike classical dynamic linking (e.g., \texttt{ld}), each module owns its linear memory and shared linear memories must be declared explicitly.

Instantiation transforms a \ac{Wasm} module into a \ac{Wasm} \textit{module instance} by allocating and initializing its linear memory, tables, and globals accordingly.
Since each instantiation is independent, a single compiled module can yield multiple concurrent instances with disjoint state.

Finally, \ac{Wasm} code is \textbf{executed} through interpretation, \ac{AOT} or \ac{JIT} compilation depending on \ac{VM} capabilities.
The host environment invokes exported functions through the \ac{VM}'s embedder \ac{API}, with module instances retaining state across successive calls~\cite{WebAssemblyCoreSpecification1}.
Execution either returns normally or produces a \textit{trap}, an error that aborts the current invocation and is returned to the host environment.

\subsubsection{Threading}
\label{subsubsection:threading}

ARINC 653 is inherently multi-threaded.
Every useful ARINC 653 partition requires at least two threads, the start process\footnote{Recall from \autoref{sec:background:arinc_653} that an ARINC 653 partition's processes, like threads, share the same address space} and at least one (a)periodic process to be scheduled in \texttt{NORMAL\_MODE}.
Further, intra-partition communication requires shared memory between a partition's processes.
Hence, basic multi-threading primitives (i.e. atomic operations for synchronization) are required.

By design, \ac{Wasm} is single-threaded: the original \ac{MVP}, version 2 and the draft for version 3\footnote{As of the 14th August 2026, version 3 does not include the threads proposal.} all permit only a single concurrent thread per \ac{Wasm} store, without shared objects between stores~\cite{WebAssemblyCoreSpecification1, WebAssemblyCoreSpecification2, watt2025concurrency}.
This restriction stems from \ac{Wasm}'s origin as a web technology built alongside JavaScript, which historically relied on an event-loop concurrency as opposed to true multi-threading.
To enable binary data sharing between JavaScript worker threads, the \texttt{SharedArrayBuffer} was eventually introduced~\cite{mdn_sharedarraybuffer_security}.

The Threads proposal lifts the single-threading restriction by employing a shared linear memory together with various atomic instructions~\cite{weakening_wasm, wasm_proposal_threads}.
On the web, it requires the \lstinline|SharedArrayBuffer|.
Outside of the web, the choice of an adequate backing storage for shared linear memories is left up to the \ac{Wasm} \acp{VM}.
At the time of writing, this proposal is in phase 4, i.e. implementations exist but it is not yet standardized.

A shared address space, as provided by the Threads proposal is a fundamental property for existing threading models such as Linux threads  or ARINC~653 processes (\autoref{tab:arinc653_partition_vs_process}).
However, enabling concurrent execution is not merely a matter of sharing memory.

Compiler toolchains targeting \ac{Wasm} typically allocate a second stack --- the so-called \textit{linear stack} --- inside the linear memory to support pointers to arbitrary elements on the stack.
When multiple \ac{Wasm} module instances share a common linear memory, these per-instance linear stacks overlap, resulting in data corruption\footnote{Pairing the \ac{PIC} relocation format for compilation with dynamic initialization of individual base addresses per linear stack solves this. However \ac{PIC} remains experimental.~\cite{wasm_tool_conventions_dynamic_linking}}.

Furthermore, the current Threads proposal only permits memories to be shared; globals, tables, and function references remain instance-local~\cite{watt2025concurrency}.
This so called instance-per-thread scheme imposes several limitations, among which is the lack of \ac{TLS} separate from the shared linear memory.
\ac{TLS}, however, can be implemented via embedding it into the linear memory, as done for this paper and explained in \autoref{subsec:init_per_process_storage}.

The \textit{Shared-Everything Threads Proposal}~\cite{wasm_proposal_shared_everything_threads} addresses aforementioned limitations.
It complements the Threads proposal, by allowing globals, tables and other \ac{Wasm} objects to be shared between threads aswell.
Also, it allows some of these objects to be explicitly marked as thread-local.
However, the shared-everything threads proposal is still a phase 1 proposal, i.e. lacking a complete specification and with ongoing design and implementation efforts.

In summary, many uncertainties around \ac{Wasm} threading remain: the initial Threads proposal is operational, but changes in the form of stable \acp{ABI}, improved toolchain configuration, or the forthcoming \textit{shared-everything threads} will eventually land.

\section{Requirements}
\label{sec:requirements}

Combining the ARINC~653 \ac{OS} \ac{API} with the \ac{Wasm} compilation target opens a vast solution space.
In order to facilitate our effort, we define requirements, guiding the subsequent planning and implementation phases.
On the system level, the following three requirements suffice:

\medskip

\requirement{SYS-1}{ARINC~653 compliant partition/application C code shall be compilable to \ac{Wasm} code.}
\requirement{SYS-2}{ARINC~653 \ac{Wasm} code shall be executable within an ARINC~653 compliant \ac{OS}'s partition.}
\requirement{SYS-3}{ARINC~653 \ac{Wasm} code shall be forwards compatible to \ac{Wasm} code execution environments provided in other ARINC~653 \acp{OS}.\footnote{confined however by the respective \ac{OS}' limitations, e.g. maximum number of processes per partition}}

The first two requirements capture the primary processes required for creation and execution of \ac{Wasm} code while relying on existing ARINC~653 compliant \acp{OS}, namely:
Compilation to \ac{Wasm} [\nameref{req:SYS-1}] and subsequent execution thereof [\nameref{req:SYS-2}].
As described in \autoref{sec:background:arinc_653}, ARINC~653 aims to provide compatibility, while actually only achieving portability.
To improve upon this, we require the system to be forwards compatible with future \ac{Wasm} code execution environments executed on other ARINC~653 \acp{OS} [\nameref{req:SYS-3}].
To comply with these system requirements, we consider the following high-level software requirements (brackets implicate traceability to parent requirements):

\medskip

\requirement{HL-1}{An ARINC~653 \ac{Wasm} code execution environment shall be provided. [\nameref{req:SYS-2}]}
\requirement{HL-2}{The ARINC~653 \ac{Wasm} code execution environment itself shall be executable inside of an ARINC~653 compliant OS. [\nameref{req:SYS-2}]}
\requirement{HL-3}{The \ac{Wasm} code execution environment shall comply to the ARINC~653 API specification. [\nameref{req:SYS-2}]}
\requirement{HL-4}{The ARINC~653 interface exposed to the \ac{Wasm} execution environment shall be unambiguous. [\nameref{req:SYS-1}, \nameref{req:SYS-3}]}
\requirement{HL-5}{The compilation of partition/application C code to \ac{Wasm} shall remain independent of the target \ac{OS}. [\nameref{req:SYS-1}, \nameref{req:SYS-3}]}

Foremost, these high-level requirements introduce the \textit{\ac{Wasm} code execution environment} [\nameref{req:HL-1}], which exists between the host side [\nameref{req:HL-2}] and guest side [\nameref{req:HL-3}] ARINC~653 environment.
One of ARINC~653's defects previously discussed is its ambiguity, for example in the bit width of index types and the order of struct fields.
To assure object code\footnote{in the form of \ac{Wasm}} level compatibility, we require the absence of such ambiguity [\nameref{req:HL-4}].
Lastly, we require compilation to remain independent of the target \ac{OS}, keeping application developers at distance of the target \ac{OS} helps to avoid tailoring which ultimately hinders compatibility [\nameref{req:HL-5}].

\section{Implementation}
\label{sec:implementation}

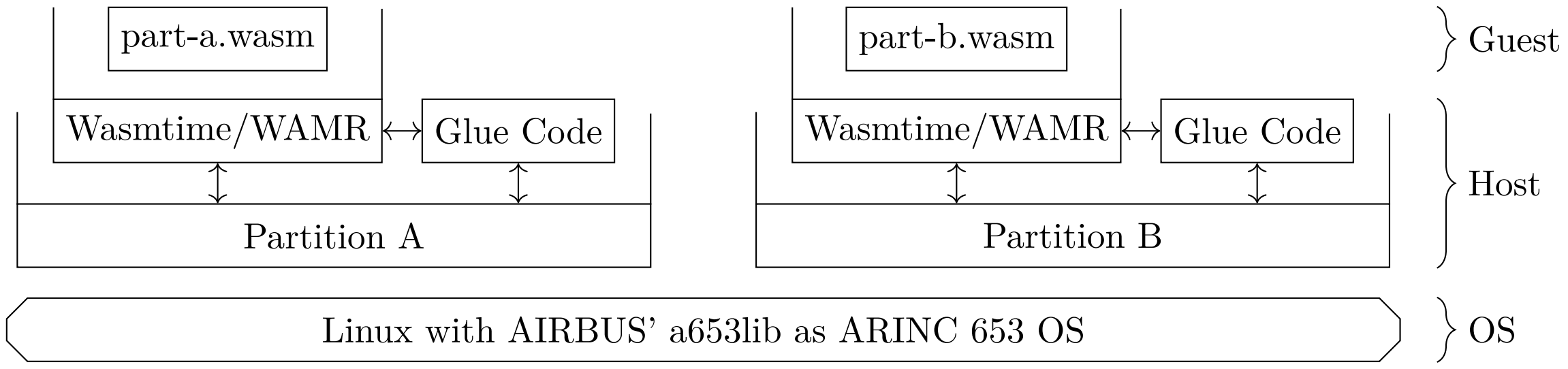
\begin{figure}
    \centering
    \begin{tikzpicture}[
    real partition/.style = {partition, minimum width = 17.5em}
]



    \begin{scope}[node distance=0.75em]
        \node[exec env, minimum width=38.5em, label=center:Linux with AIRBUS' a653lib as ARINC 653 \ac{OS}] (os) {};
        \node[real partition, above left = 2em and 0em of os, anchor=west] (partition1) {Partition A};
        \node[real partition, above right = 2em and 0em of os, anchor=east] (partition2) {Partition B};
        \node[partition, artifact, above left = 2em and -1em of partition1, anchor=west] (interpreter1) {Wasmtime/WAMR};
        \node[artifact, above right = 2em and -1em of partition1, anchor=east] (glue1) {Glue Code};
        \node[artifact, above = of interpreter1] (wasm1) {part-a.wasm};
        \node[partition, artifact, above left = 2em and -1em of partition2, anchor=west] (interpreter2) {Wasmtime/WAMR};
        \node[artifact, above right = 2em and -1em of partition2, anchor=east] (glue2) {Glue Code};
        \node[artifact, above = of interpreter2] (wasm2) {part-b.wasm};
        
        \draw[<->] (glue1) -- (interpreter1);
        \draw[<->] (glue2) -- (interpreter2);
        \draw[<->] (glue1.south) -- (partition1.north -| glue1.south);
        \draw[<->] (glue2.south) -- (partition2.north -| glue2.south);
        \draw[<->] (interpreter1.south) -- (partition1.north -| interpreter1.south);
        \draw[<->] (interpreter2.south) -- (partition2.north -| interpreter2.south);
        
        \coordinate (labelheight) at ($(os.east)+(0,0)$);
        \coordinate (guestenvbot) at (labelheight  |- wasm2.south);
        \coordinate (guestenvtop) at (labelheight  |- wasm2.north);
        \coordinate (hostenvbot) at (labelheight  |- partition2.south);
        \coordinate (hostenvtop) at (labelheight  |- glue2.north);
        \coordinate (osbot) at (labelheight  |- os.south);
        \coordinate (ostop) at (labelheight  |- os.north);
        
        \draw[decorate,decoration={brace,amplitude=5pt,mirror,raise=1em}](guestenvbot) -- (guestenvtop)
        node[pos=0.5,right=1.5em,black]{Guest};
        \draw[decorate,decoration={brace,amplitude=5pt,mirror,raise=1em}] (hostenvbot) -- (hostenvtop)
        node[pos=0.5,right=1.5em,black]{Host};
        \draw[decorate,decoration={brace,amplitude=5pt,mirror,raise=1em}] (osbot) -- (ostop)
        node[pos=0.5,right=1.5em,black]{\ac{OS}};
    \end{scope}
 \end{tikzpicture}
    \caption{Demonstrator architecture}
    \label{fig:demo_architecture}
\end{figure}

Guided by the requirements introduced in \autoref{sec:requirements}, a demonstration was implemented.
To enable unobstructed reproduction of the demonstration, we opted to only rely on open-source components.
Namely AIRBUS' ARINC 653 shell based on Linux\footnote{\url{https://github.com/airbus/a653lib}} is used as the \ac{OS}.
Meanwhile both Wasmtime and \ac{WAMR} are used to demonstrate independence of any specific \ac{Wasm} \ac{VM}.
\autoref{fig:demo_architecture} illustrates the chosen embedding of \ac{Wasm} \acp{VM} into the ARINC~653 host environment, using environment definitions from \autoref{fig:tech_stack}.
In particular, the use of a \ac{Wasm} \ac{VM} within the partitions remains an implementation detail of the partition.
Partitioning is provided by the \ac{OS}' partitioning mechanism, albeit enhanced with the additional isolation provided by the \ac{Wasm} \acs{VM}'s sandbox.
The application resident in a partition is written in C and compiled into bytecode forming a single \ac{Wasm} module \footnote{Ada and Rust are also feasible due to their C-\ac{ABI} compatible \ac{FFI}, but are not part of the initial implementation.}.
In-between that application and the \ac{OS}, a shim is installed comprising the \ac{Wasm} \ac{VM} and glue code for translation.
We sketched this concept for the first time in \cite[Fig. 5]{10311207}.

Note that we explicitly choose to implement the host environment as a middleware that forwards \ac{APEX} calls from the guest environment to the \ac{OS}.




To implement the architecture sketched in \autoref{fig:demo_architecture}, five major roadblocks must be resolved.
The following sections discuss these in detail.

\subsection{Tying up the ARINC~653 \texorpdfstring{\ac{APEX}}{APEX} Interfaces}
\label{sec:arinc653_wasm_headers}

To even compile an ARINC~653 program, one has to pick on various implementation dependent aspects of ARINC~653.
Usually, this is already done by the \ac{OS} vendor.
However, as \ac{OS} independence is required [\nameref{req:HL-5}] and all ambiguities must be resolved [\nameref{req:HL-4}], this subsection outlines \textit{our} choices and corresponding rationale (see \autoref{apx:rationale} for all design decisions also unrelated to interface design).

ARINC~653 example header files for Part 1 and Part 2 can be downloaded from the SAE website\footnote{\label{footnote:arinc_headers}\url{https://www.aviation-ia.com/support-files/arinc653h}, \url{https://www.aviation-ia.com/support-files/arinc653p2h}}.
However, these constitute only one of multiple ARINC~653 compliant interpretations, and thus their use is not compulsory.
For the scope of this paper, we base our work on the header file for part 1 distributed by SAE/ARINC.
These header files are not provided under any license, thus instead of providing the final header file, we provide an AWK script to insert our changes\footnote{\url{https://github.com/DLR-FT/arinc653-wasm/blob/main/scripts/process-arinc-header.awk}}.
To the best of our knowledge, all of our changes remain in the realm of implementation defined behavior and/or are purely syntactical.
Hence we assume standard compliance to remain intact despite our changes, which amount to the following:

\begin{enumerate}
    \item \implchoice{For the implementation dependent \lstinline|#define| macros, use \lstinline|#ifndef| guards to avoid re-definition of all \lstinline|SYSTEM_LIMIT_*| values}{This enables one canonical ARINC~653 \ac{Wasm} header file, that without being changed, supports systems with varying \lstinline|SYSTEM_LIMIT_*| values}
    \item \implchoice{Prefix all ARINC~653 function declarations with the \lstinline|WASM_IMPORT_MODULE| macro}{When compiling to a \ac{Wasm} target, this macro expands to compiler attributes instructing the linker to import these functions at runtime.
        Consequently, before running the resulting \ac{Wasm} bytecode, the host environment must provide implementations for all ARINC~653 functions used in the code.
        For non \ac{Wasm} targets, the macro expands to an empty string, retaining the ability to use the same header file}
    \item \implchoice{Replace void pointers to functions with actual function pointers}{The original ARINC~653 API is specified to use void pointers for function pointers.
        While this may work with \ac{Wasm} at the time of writing\footnote{The compiler represents these function pointers as table indices, refer to \autoref{subsubsec:reference_types}.}, it is imprecise and hides the function's signature, reducing the compiler's ability to verify the code.
        Therefore, we are uncertain about future support for void pointers as function pointers for \ac{Wasm}}
    \item \implchoice{Where integer bit width is implementation dependent, always pick the biggest option}{The header offered for download is not parse-able C code, as it contains a handful of time the string \lstinline|<an APEX integer type>| instead of a type.
        Each occurrence of this is for an \textit{id} type referring to an \ac{OS} managed resource, such as a partition or sampling port.
        Thus, all of these are always initialized by the \ac{OS}.
        From a data retention perspective, using the 64 bit integers instead of 32 bit integers for the \textit{id} types is sound; even if the \ac{OS} implements the type as a 32 bit integer, all possible valid values of that can also be stored in a 64 bit integer.
        Data access wise, refer to \autoref{sec:abi_translation}}
\end{enumerate}

While our changes to the ARINC~653 header file are focused on \ac{Wasm}, we maintain utility of the ARINC~653 header file independent of \ac{Wasm} as compilation target.
To this end, we would like to see more of an open ecosystem to form around ARINC~653, enabled both legally (by a license for the already published header files) and technically, by removing some of the cruft as the reference to the ANSI C standard (which has been withdrawn in favor of \cite{ISO9899}), and relying on \lstinline|stdint.h| for guaranteed integer bit widths.


\subsection{C-ABI-Lens: Accessing Data in Foreign ABI Representation}
\label{sec:abi_translation}

For this subsection, we assume a code execution environment according to \nameref{req:HL-1} is provided (which is done in \autoref{sec:porting-the-arinc653-interfaces}).
Ultimately, when an ARINC~653 function is called in the guest environment, it must be handled in the host environment.
More specifically, the control-flow transfers from the guest environment to the host environment, which is itself a partition's process in an ARINC~653 compliant \ac{OS} [\nameref{req:HL-2}].
As outlined in \autoref{sec:implementation}, this host environment shall act as a mediator and therefore it has to relay the ARINC~653 function call to the \ac{OS}.
This subsection addresses the implications of this process in the light of \ac{ABI} differences.

ARINC~653 heavily relies on pass-by-reference for passing data in and out of functions.
In fact, all ARINC~653 Part 1 functions return \lstinline|void|, and even the ubiquitous \lstinline|RETURN_CODE_TYPE| which indicates success/error is exclusively passed by reference.
Furthermore, structures such as \lstinline|PROCESS_STATUS_TYPE| are passed by reference too.

Passing primitive types for which an equivalent exists in \ac{Wasm} is very straight-forward.
However, while pointers are represented as integer primitives in \ac{Wasm}, the actual layout of the data pointed to may deviate from the host's expectation of its memory layout; thus two complications arise.

On the one hand, ARINC~653 does not specify the order of fields, thus the ARINC~653 \ac{API} in the guest environment may have struct fields in a different order than that of the host environment's \ac{OS}.
Therefore, struct translation must take place.

On the other hand, alignment related padding and different endianness\footnote{Many avionics systems run on embedded PowerPC architectures, which defaults to big-endian; and often even allows endianness to vary per memory page, while \ac{Wasm} is strictly little-endian.} complicate this further.
Even when the host environment has access to the specific ARINC~653 header files that the guest-code was compiled with (such as the outcome of \autoref{sec:arinc653_wasm_headers}), accessing the data is still non-trivial due to \ac{ABI} specific layout and representation of data in memory.
The very same instance (value-wise) of a struct might have differing size and vastly different in-memory bit-wise representation when stored in host vs. guest environment.

To illustrate this, consider a call to the \lstinline|GET_PROCESS_STATUS|\textsubscript{Guest} function declared in \autoref{lst:process_status_type}.
It is passed a process id by value and two pointers, to the process status struct, and to a return code integer.
Then, the \ac{OS} propagates the struct fields with fresh data, and writes to the return code, indicating success or failure.
In fact, there are two interpretations of the struct: \lstinline|PROCESS_STATUS_TYPE|\textsubscript{Guest} (represented as \ac{Wasm} would layout the struct in its linear memory) and \lstinline|PROCESS_STATUS_TYPE|\textsubscript{Host/\ac{OS}} (represented according to the current native target, e.g. x86\_64-linux).

\begin{lstlisting}[
    style=C,
    label=lst:process_status_type,
    float,caption={Example \ac{APEX} struct and function definition.}
]
typedef
   struct {
      SYSTEM_TIME_TYPE        DEADLINE_TIME;
      PRIORITY_TYPE           CURRENT_PRIORITY;
      PROCESS_STATE_TYPE      PROCESS_STATE;
      PROCESS_ATTRIBUTE_TYPE  ATTRIBUTES;
   } PROCESS_STATUS_TYPE;

extern void GET_PROCESS_STATUS (
       /*in */ PROCESS_ID_TYPE          PROCESS_ID,
       /*out*/ PROCESS_STATUS_TYPE      *PROCESS_STATUS,
       /*out*/ RETURN_CODE_TYPE         *RETURN_CODE );
\end{lstlisting}

\begin{lstlisting}[
    style=C,
    label=lst:process_status_type_c_abi_lens,
    float,caption={Getter/setter generated by \texttt{c-abi-lens} for the struct from \autoref{lst:process_status_type}}
]
/************************** PROCESS_STATUS_TYPE ***************************/
uintptr_t cal_sizeof__PROCESS_STATUS_TYPE(void);

/******************* PROCESS_STATUS_TYPE.DEADLINE_TIME ********************/
uintptr_t cal_sizeof__PROCESS_STATUS_TYPE__DEADLINE_TIME(void);
int64_t cal_get__PROCESS_STATUS_TYPE__DEADLINE_TIME(void *struct_base_addr);
void cal_set__PROCESS_STATUS_TYPE__DEADLINE_TIME(void *struct_base_addr,
                                                 int64_t value);
uintptr_t cal_offsetof__PROCESS_STATUS_TYPE__DEADLINE_TIME(void);
// ...
\end{lstlisting}

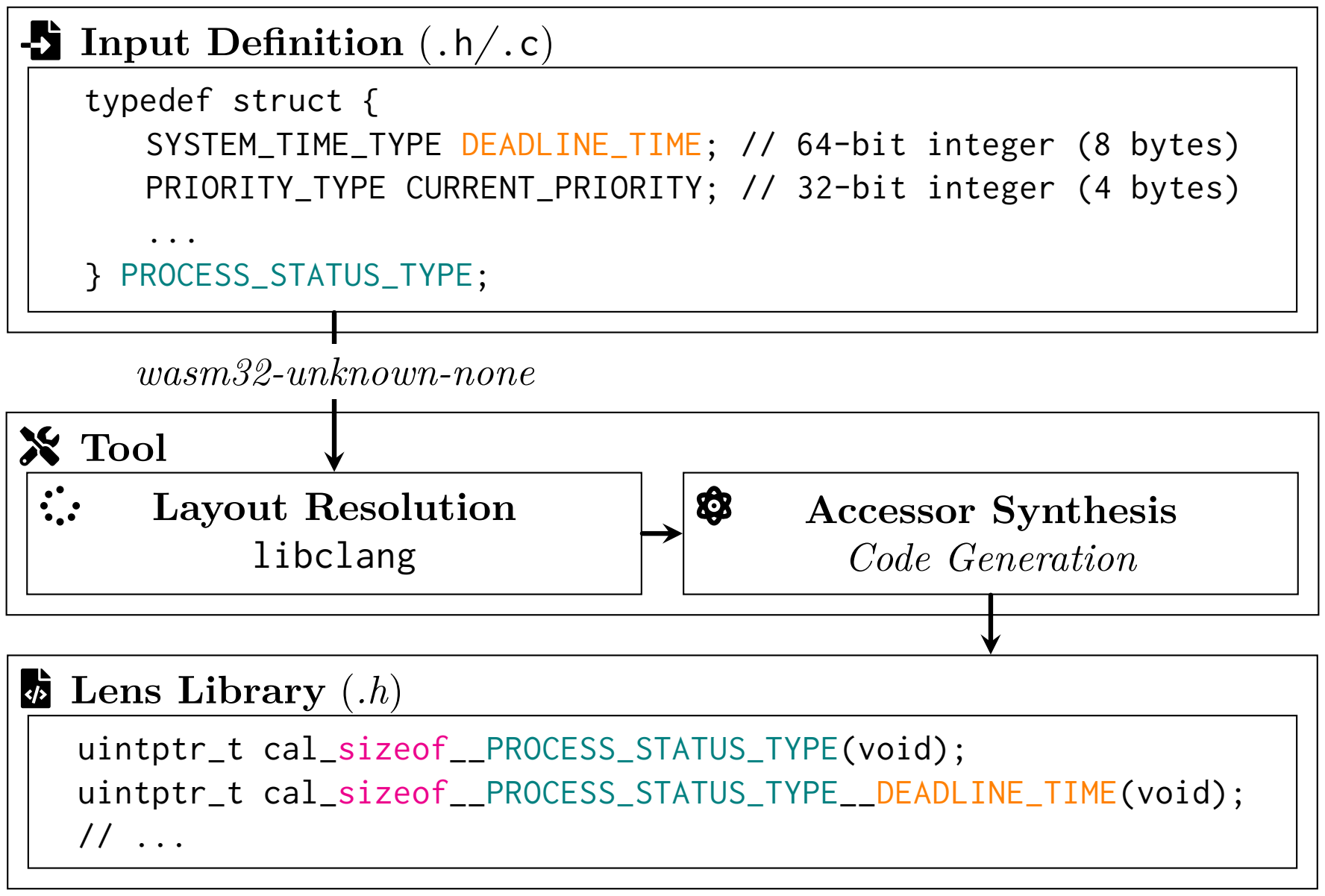
\begin{figure}[H]
    \centering
    \begin{tikzpicture}[
    box/.style = {draw, align = center, inner ysep = 0.5em, inner xsep = 0.5em},
    code/.style = {box, minimum width=32.0em, align=left, font=\ttfamily\small}
]
	\node[box, minimum width=15.5em] (layout_res) {\textbf{Layout Resolution}\\\texttt{libclang}};
	\node[anchor = north west] at (layout_res.north west) {\faSpinner};

	\node[fit=(layout_res.north west) (layout_res.south east), inner sep=0, right = 1.0em of layout_res] (synth_w) {};
	\node[align = center] (synth_text) at (synth_w) {\textbf{Accessor Synthesis}\\\textit{Code Generation}};
	\node[box, fit=(synth_w) (synth_text), inner sep=-0.4pt] (synth_box) {};
	\node[anchor = north west] at (synth_box.north west) {\faAtom};

	\node[box, fit=(layout_res) (synth_box), yshift=0.5em, inner ysep = 1.0em] (tool) {};
	\node[anchor = north west] at (tool.north west) {\faTools\hspace{0.5em}\textbf{Tool}};

	\node[code, above = 2.5em of tool.north] (code_in) {
		typedef \textbf{struct} \{ \\
		\hspace{2em}SYSTEM\_TIME\_TYPE \textcolor{orange}{DEADLINE\_TIME}; // 64-bit integer (8 bytes) \\
		\hspace{2em}PRIORITY\_TYPE CURRENT\_PRIORITY; // 32-bit integer (4 bytes) \\
		\hspace{2em}... \\
		\} \textcolor{teal}{PROCESS\_STATUS\_TYPE};
	};
	\node[box, fit=(code_in) , yshift=0.5em, inner ysep = 1.0em] (code_in_box) {};
	\node[anchor = north west] at (code_in_box.north west) {\faFileImport\hspace{0.5em}\textbf{Input Definition} (\texttt{.h}\textit{/}\texttt{.c})};

	\node[code, below = 2.5em of tool.south] (code_out) {
		\textbf{uintptr\_t} cal\_\textcolor{magenta}{sizeof}\_\_\textcolor{teal}{PROCESS\_STATUS\_TYPE}(void); \\
		\textbf{uintptr\_t} cal\_\textcolor{magenta}{sizeof}\_\_\textcolor{teal}{PROCESS\_STATUS\_TYPE}\_\_\textcolor{orange}{DEADLINE\_TIME}(void); \\
		// ...
	};
	\node[box, fit=(code_out) , yshift=0.5em, inner ysep = 1.0em] (code_out_box) {};
	\node[anchor = north west] at (code_out_box.north west) {\faFileCode\hspace{0.5em}\textbf{Lens Library} (\textit{.h})};

	\draw[very thick, line cap=round, -stealth] (code_in.south -| layout_res.north) -- (layout_res.north) node[pos=0.37, fill=white] {\textit{wasm32-unknown-none}};
	\draw[very thick, line cap=round, -stealth] (layout_res.east) -- (synth_box.west);
	\draw[very thick, line cap=round, -stealth] (synth_box.south) -- (code_out_box.north -| synth_box.south);
 \end{tikzpicture}
    \caption{Workflow of \texttt{c-abi-lens}. A header file (e.g. \autoref{lst:process_status_type}) is processed according to a specific target \ac{ABI} (i.e. \textit{wasm32-unknown-none}), deriving the memory layout of all contained structure types. Then, functions are generated to read/write fields and reflect on properties such as size for structs in that target \ac{ABI}'s memory layout (q.v. \autoref{lst:process_status_type_c_abi_lens})}
    \label{fig:lens_flow}
\end{figure}

\begin{figure}[ht]
    \centering
    \begin{tikzpicture}[
	hex/.style = {draw, inner sep=0.23em},
    box/.style = {draw, align = center, inner ysep = 1em, inner xsep = 1.5em},
    code/.style = {box, minimum width=30.75em, inner sep=0.5em, align=left, font=\ttfamily\small}
]
	\node[hex, fill=gray!50] (i0c) {\texttt{0x11}};
	\node[hex, right = -0.4pt of i0c, fill=gray!50] (i1c) {\texttt{0x22}};
	\node[hex, right = -0.4pt of i1c, fill=gray!50] (i2c) {\texttt{0x33}};
	\node[hex, right = -0.4pt of i2c, fill=gray!50] (i3c) {\texttt{0x44}};
	\node[hex, right = -0.4pt of i3c, fill=gray!50] (i4c) {\texttt{0x55}};
	\node[hex, right = -0.4pt of i4c, fill=gray!50] (i5c) {\texttt{0x66}};
	\node[hex, right = -0.4pt of i5c, fill=gray!50] (i6c) {\texttt{0x77}};
	\node[hex, right = -0.4pt of i6c, fill=gray!50] (i7c) {\texttt{0x88}};
	\node[hex, right = -0.4pt of i7c, fill=orange!50] (i8c) {\texttt{0x00}};
	\node[hex, right = -0.4pt of i8c, fill=orange!50] (i9c) {\texttt{0x00}};
	\node[hex, right = -0.4pt of i9c, fill=orange!50] (i10c) {\texttt{0x00}};
	\node[hex, right = -0.4pt of i10c, fill=orange!50] (i11c) {\texttt{0x64}};

	\node[draw, inner sep=-0.2pt, fit=(i0c) (i7c), below=-0.4pt of i0c.south west, anchor=north west] (l0) {};
	\node at (l0) {\texttt{DEADLINE\_TIME}};
	\node[draw, inner sep=-0.2pt, fit=(i8c) (i11c), below=-0.4pt of i8c.south west, anchor=north west] (l0) {};
	\node at (l0) {\texttt{CURRENT\_PRIORITY}};

	\path (i8c.north west) -- ++(0,2em) coordinate (lens_l);
	\path (i11c.north east) -- ++(0,2em) coordinate (lens_r);

	\draw[line cap=round, thick] (lens_l) .. controls +(1, 0.3) and +(-1, 0.3).. (lens_r)
		coordinate[midway] (lens_05t);
	\draw[line cap=round, thick] (lens_l) .. controls +(1, -0.3) and +(-1, -0.3).. (lens_r)
	    coordinate[pos=0.25] (lens_025) coordinate[pos=0.75] (lens_075);

	\draw[line cap=round, thick] (i8c.north west) -- (lens_025);
	\draw[line cap=round, thick] (i11c.north east) -- (lens_075);

	\draw[line cap=round, thick] (i0c.north west) ++(0,2em) -- (lens_l) node[midway, fill=white] {OFFSET = 8};
	\path (i10c.north west) -- ++(0,2em) node[pos=1,yshift=-1pt] {\texttt{memcpy}};

	\node[hex, above = 5.5em of i8c, fill=lime!50] (i8ct) {\texttt{0x64}};
	\node[hex, right = -0.4pt of i8ct, fill=lime!50] (i9ct) {\texttt{0x00}};
	\node[hex, right = -0.4pt of i9ct, fill=lime!50] (i10ct) {\texttt{0x00}};
	\node[hex, right = -0.4pt of i10ct, fill=lime!50] (i11ct) {\texttt{0x00}};
	\node[hex, inner sep=-0.2pt, fit=(i8ct) (i11ct), above=-0.4pt of i8ct.north west, anchor=south west] (l0t) {};
	\node at (l0t) {\texttt{PRIORITY\_TYPE}};

	\draw[line cap=round, stealth-stealth, thick] (lens_05t) -- (i10ct.south west) node[midway, fill=white] {\texttt{bswap\_32}}; 

	\node[code, below = 2em of i6c.south west] (code) {
		typedef struct \{ \\
		\hspace{2em}SYSTEM\_TIME\_TYPE \textcolor{gray!70!black}{DEADLINE\_TIME}; // 64-bit integer (8 bytes) \\
		\hspace{2em}PRIORITY\_TYPE \textcolor{orange}{CURRENT\_PRIORITY}; // 32-bit integer (4 bytes) \\
		\hspace{2em}... \\
		\} PROCESS\_STATUS\_TYPE;
	};

	\path (code.east) -- ++(0.5em,0) node[pos=1, right] {Struct Definition};
	\path (i11c.south east) -- ++(0.5em,0) node[pos=1, right] {Foreign ABI Layout};
	\path (lens_r) -- ++(0.5em,0) node[pos=1, right] {Lens Library (\texttt{.h})};
	\path (i11ct.north east) -- ++(0.5em,0) node[pos=1, right] {Host ABI};

 \end{tikzpicture}
    \caption{Calling a \lstinline|cal_get__| function (q.v. \autoref{lst:process_status_type_c_abi_lens}) generated by \texttt{c-abi-lens}. The access functions receive a pointer to the start address of a struct. The relevant bytes (defined by the respective field's \lstinline|offsetof| in the target \ac{ABI}'s memory layout) are copied into the target type (i.e. \lstinline|int32_t|). If the foreign \ac{ABI}'s endianness differs from the native one \lstinline|bswap_*| is used to convert it, before returning the final value.}
    \label{fig:lens_call}
\end{figure}
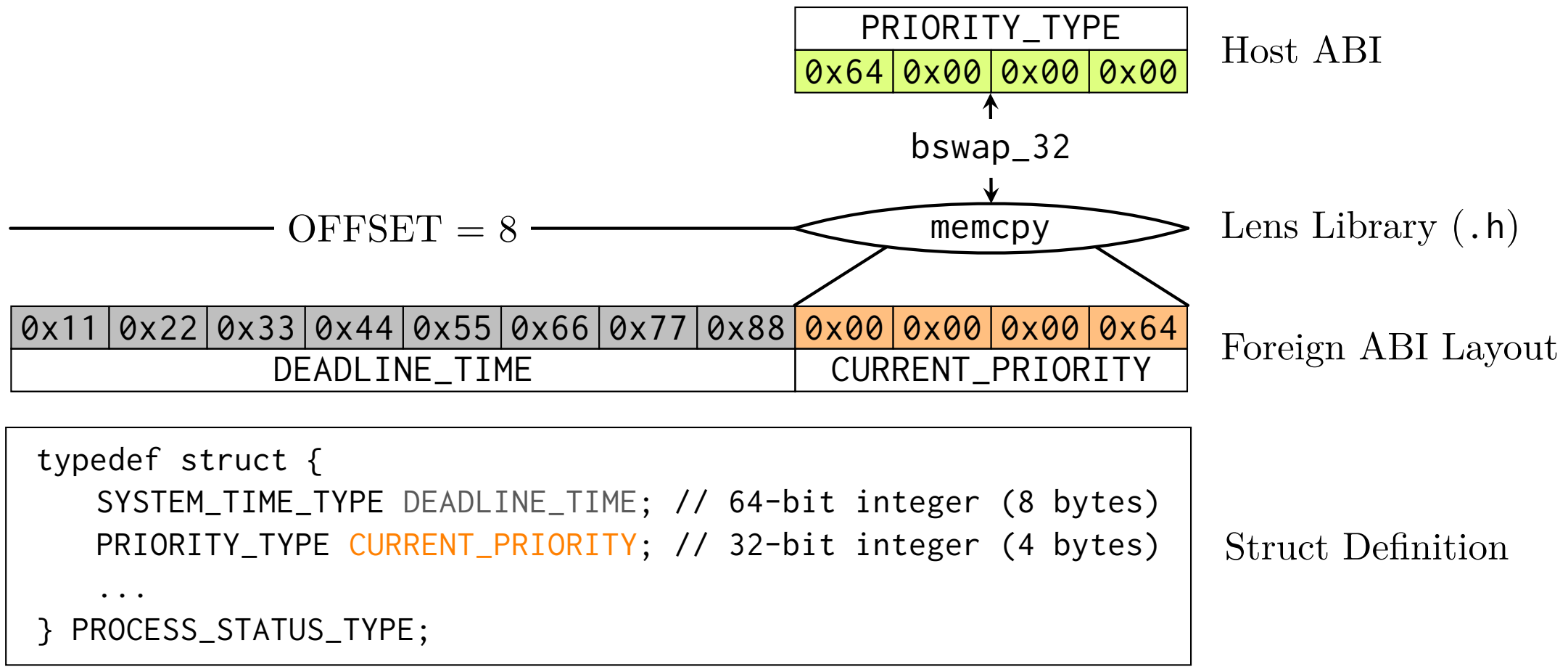

Due to the potential difference of in-memory representations, code is required to get and set data in instances of \lstinline|GET_PROCESS_STATUS|\textsubscript{Guest} residing in the \ac{Wasm}'s linear memory.
To alleviate this problem, we built \texttt{c-abi-lens}\footnote{\url{https://github.com/DLR-FT/arinc653-wasm/tree/main/pkgs/c-abi-lens}}, a code-generator to derive thin data access libraries for C structures represented in a foreign \ac{ABI}.
Upon execution, \texttt{c-abi-lens} is passed a header file (e.g. ARINC653-wasm.h) and a target architecture (e.g. \textit{wasm32-unknown-none}), as illustrated in \autoref{fig:lens_flow}.
It then generates C code with access functions for structures of that target architecture's in-memory representation.
These functions (see \autoref{fig:lens_call} and \autoref{lst:process_status_type_c_abi_lens} for an example) then allow to read and write each field and to reserve memory for an entire instance of the struct.
Crucially, while the generated code is architecture specific regarding the in-memory representation of structs to be accessed (\textit{wasm32-unknown-none} in our example), it is agnostic to which architecture the generated code is compiled to.

Equipped with the \texttt{c-abi-lens} code generator, relaying ARINC~653 calls now becomes feasible.
Sticking to the example of \lstinline|GET_PROCESS_STATUS|, the following series of events takes place.

\begin{enumerate}
    \item Guest environment calls \lstinline|GET_PROCESS_STATUS|\textsubscript{Host}, sending control flow to the host environment.
    \item Host environment allocates an instance of \lstinline|PROCESS_STATUS_TYPE|\textsubscript{Host/\ac{OS}} and an instance of \lstinline|RETURN_CODE_TYPE|\textsubscript{Host/\ac{OS}}.
    \item Host environment calls \lstinline|GET_PROCESS_STATUS|\textsubscript{\ac{OS}} with the process id provided by the guest but pointers to the two instances from step 2.
        Then, control flows to the \ac{OS}, and subsequently back to the host environment.
    \item Host environment derives the base address to the \lstinline|PROCESS_STATUS_TYPE|\textsubscript{Guest} and \lstinline|RETURN_CODE_TYPE|\textsubscript{Guest} inside the \ac{Wasm}'s linear memory.
        These are computed as the sum of the linear memory's start address and respective pointer passed from guest to host in step 1.
    \item For each field in \lstinline|PROCESS_STATUS_TYPE|\textsubscript{Guest}, the appropriate \lstinline|cal_set__PROCESS_STATUS_TYPE__*| function is used to set the field in \lstinline|PROCESS_STATUS_TYPE|\textsubscript{Guest} to the value found in the \lstinline|GET_PROCESS_STATUS|\textsubscript{Host/\ac{OS}} from step 2.
    \item The value of \lstinline|RETURN_CODE_TYPE|\textsubscript{Host} is written to the address of \lstinline|RETURN_CODE_TYPE|\textsubscript{Guest} (which was calculate in step 4 from the pointer passed in step 1).
    \item Host environment continues execution of the guest environment, sending control flow to the guest environment (via virtue of the \ac{VM}).
\end{enumerate}

This schema universally enables the pass-by-reference nature prescribed by ARINC~653, despite host and guest environment laying out the same structures differently in memory.

\subsection{Handling of function pointers}
\label{sec:handling_of_function_pointers}

As explained in \autoref{subsubsec:reference_types}, \ac{Wasm} handles function pointers/references differently than other forms of data.
In order to comply with the ARINC~653 \ac{API}~[\nameref{req:HL-3}] and correctly implement the glue code in the host environment~[\nameref{req:HL-2}], additional work is needed.
This section describes the problem at hand and highlights possible solutions.

In the ARINC~653 \ac{API}, two functions are concerned with the passing of function pointers: \lstinline|CREATE_PROCESS| and \lstinline|CREATE_ERROR_HANDLER|.
Both normally expose a function within the partition for invocation by the \ac{OS}.
When using \ac{Wasm}, this causes a problem:
as the representation of \ac{Wasm} functions is \ac{Wasm}-\ac{VM}-specific and the \ac{OS} does not have any knowledge of \ac{Wasm}, it can not call a \ac{Wasm} function directly.
Thus, function calls from the \ac{OS} must be relayed by the host environment to the guest environment.

A naive approach is the creation of one wrapper function in the host environment per function in the guest environment (1:1 mapping).
However, an ARINC~653 partition compiled to \ac{Wasm} could compose an arbitrary number of functions.
Thus, the 1:1 mapping is only feasible when knowing at compile-time of the host environment how many functions the guest environment might have, as the host environment needs to be customized based on information from the \ac{Wasm} bytecode.
Such dependence hinders software re-use and flexibility, main drivers for both ARINC~653 and our related \ac{Wasm} efforts.

More involved, but generic over any imaginable \ac{Wasm} bytecode is a single dispatch function approach.
When the guest environment passes a function pointer to the host environment via one of the two aforementioned ARINC~653 calls, the host environment denotes the context of that function pointer in a persistent data structure.
It then itself forwards the ARINC~653 call to the \ac{OS}, however replacing the \ac{Wasm} function pointer with one singular dispatch function resident in the host environment.
Upon invocation of this dispatch function by the \ac{OS}, it looks up the corresponding \ac{Wasm} function index in that data structure and initializes a new \ac{Wasm} \ac{VM} with that \ac{Wasm} function as entry point.
For process creation, the following algorithm, also depicted in \autoref{fig:create_process}, suffices:

\begin{figure}
    \centering
    \begin{tikzpicture}[
    box/.style = {draw, align = center, inner ysep = 0.5em, inner xsep = 0.5em},
    tt/.style = {font=\ttfamily\small, text height=0.5em, inner xsep = 0em},
    ttt/.style = {tt, font=\ttfamily\footnotesize, inner xsep=0.5pt, inner ysep=1.0pt},
	wbox/.style = {box, inner ysep=0.15em, inner xsep=0.7em},
	boxtt/.style = {box, tt},
	boxttl/.style = {boxtt, align=left},
	arr/.style = {-stealth },
	arr_round/.style = {arr, line cap=round},
    code/.style = {box, align=left, font=\ttfamily\small}
]
	\node[tt] (create_wasm) {CREATE\_PROCESS()};
	\node[wbox, draw=none, fit = (create_wasm)] (create_wasm_box) {};

	\node[tt, right = 1em of create_wasm] (attr_wasm_name) {NAME};
	\node[tt, below = 0em of attr_wasm_name.south west, anchor=north west] (attr_wasm_addr) {ENTRY};
	\node[tt, below = 0em of attr_wasm_addr.south west, anchor=north west] (attr_wasm_rest) {\ldots};
	\node[wbox, dashed, fit=(attr_wasm_name) (attr_wasm_addr) (attr_wasm_rest)] (attr_wasm) {};

	\node[code, right = 9.8em of attr_wasm.north east, anchor=north west] (code_wasm) {void foo(void) \{ \\
		\hspace{1em} // \ldots \\
		\}
	};
	\draw[arr] (code_wasm.west |- attr_wasm.east) -- (attr_wasm_addr.east);

	\node[ttt, right = 0.2em of attr_wasm_addr -| attr_wasm.east, fill=white, text=teal] {\&foo};

	\node[tt, below = 4.0em of create_wasm] (create_nat) {CREATE\_PROCESS()};
	\node[wbox, draw=none, fit=(create_nat)] (create_nat_box) {};
	\node[tt, right = 1em of create_nat] (attr_nat_name) {NAME};
	\node[tt, below = 0em of attr_nat_name.south west, anchor=north west, text=green!60!black] (attr_nat_addr_new) {ENTRY};
	\node[tt, left = 0em of attr_nat_addr_new, text=green!60!black] () {+};
	\node[tt, below = 0em of attr_nat_addr_new.south west, anchor=north west, text=red!90!black] (attr_nat_addr_old) {ENTRY};
	\node[tt, left = 0em of attr_nat_addr_old, text=red!90!black] () {-};
	\node[tt, below = 0em of attr_nat_addr_old.south west, anchor=north west] (attr_nat_rest) {\ldots};
	\node[wbox, dashed, fit=(attr_nat_name) (attr_nat_addr_old) (attr_nat_addr_new) (attr_nat_rest)] (attr_nat_box) {};
	\draw[arr_round, very thick] (create_wasm_box.south) -- (create_nat_box.north);

	\node[code, right = 9.8em of attr_nat_box.north east, anchor=north west] (code_nat) {void dispatch(void) \{ \\
		\hspace{1em}uint64\_t \textcolor{orange}{pid}; \\
		\hspace{1em}GET\_MY\_ID(*\textcolor{orange}{pid}); \\
		\hspace{1em}uint32\_t \textcolor{teal}{fn\_idx}; \\
		\hspace{1em}lookup\_wasm\_fn\_index(*\textcolor{teal}{fn\_idx}, \textcolor{orange}{pid}); \\
		\hspace{1em}interpreter\_start(\textcolor{teal}{fn\_idx}); \\
		\}
	};
	\draw[arr] (code_nat.west |- attr_nat_addr_new.east) -- (attr_nat_addr_new.east);

	\node[tt, right = 7.6em of attr_nat_rest, yshift=-1em] (table_addr) {addr}; 
	\node[wbox, fit=(table_addr)] (table_addr_box) {};
	\node[tt, left = 1.4em of table_addr.north west, anchor=north east] (table_id) {pid}; 
	\node[wbox, fit=(table_id.north west) (table_id.east |- table_addr.south)] (table_id_box) {};
	\node[tt, below = 0.15em of table_addr_box, text=teal] (table_wasm_addr) {\&foo};
	\node[wbox, fit=(table_addr), yshift=-1.5em] (table_wasm_addr_box) {};
	\node[tt, below = 0.15em of table_id_box, text=orange] (table_wasm_id) {2};
	\node[wbox, fit=(table_id.north west) (table_id.east |- table_addr.south), yshift=-1.5em] (table_wasm_id_box) {};
	
	\draw[arr_round] (attr_nat_addr_old.east) -- (attr_nat_addr_old.east -| table_addr.north) -- (table_addr_box.north); 
	\draw[arr, dashed] (attr_wasm.south) -- (attr_nat_box.north) node[pos=0.5] (wasm_nat_mid) {};

	\node[ttt, right = 0.2em of attr_nat_addr_new -| attr_nat_box.east, fill=white] {\&dispatch};
	\node[ttt, right = 0.2em of attr_nat_addr_old -| attr_nat_box.east, fill=white, text=teal] {\&foo};

	\node[tt, below = 8.2em of create_nat] (os653) {ARINC\ 653 OS}; 
	\node[wbox, double, outer sep=0.8pt, fit=(os653)] (os653_box) {};
	\draw[arr_round, very thick] (create_nat_box.south) -- (os653_box.north);
	\draw[arr, very thick] (os653_box.25) -- (os653_box.25 |- table_wasm_id_box.west) -- (table_wasm_id_box.west);
	\node[ttt, left = 0.8em of table_wasm_id_box, fill=white] {PROCESS\_ID = \textcolor{orange}{2}};

	\path (table_wasm_addr_box.east) -- (table_wasm_addr_box.east -| code_nat.west) node[pos=0.5] (table_code_mid) {};
	\draw[arr] (table_wasm_addr_box.east) -- (table_wasm_addr_box.east -| table_code_mid) -- (table_code_mid |- code_nat.west) -- (code_nat.west) -- ++(1em, 0);
	\node[ttt, left = 0.2em of code_nat.west, fill=white, text=teal] {\&foo};

	\draw[decorate,decoration={brace,amplitude=5pt,mirror,raise=1em}] (code_wasm.south -| code_nat.east) -- (create_wasm.north -| code_nat.east) 
        node[pos=0.5,right=1.5em,black]{Guest};
	\draw[decorate,decoration={brace,amplitude=5pt,mirror,raise=1em}] (code_nat.south east) -- (code_nat.north east)
        node[pos=0.5,right=1.5em,black]{Host};
	\draw[decorate,decoration={brace,amplitude=5pt,mirror,raise=1em}] (os653.south -| code_nat.east) -- (os653.north -| code_nat.east)
        node[pos=0.5,right=1.5em,black]{OS};
 \end{tikzpicture}
    \caption{Relaying of \lstinline|CREATE_PROCESS| calls from guest-, via host environment, to the \ac{OS}. \ac{Wasm} function pointers (which actually are indices into a \ac{Wasm} Table) can not directly be called from the host environment. Therefore the host environment retains them in a lookup table, while instead annunciated a singular \lstinline|dispatch()| function from the host environment to the \ac{OS}. Later, upon invocation by the \ac{OS}, that \lstinline|dispatch()| function fetches the corresponding \ac{Wasm} function index, and starts a \ac{Wasm} \ac{VM} using that index as its entry point.}
    \label{fig:create_process}
\end{figure}
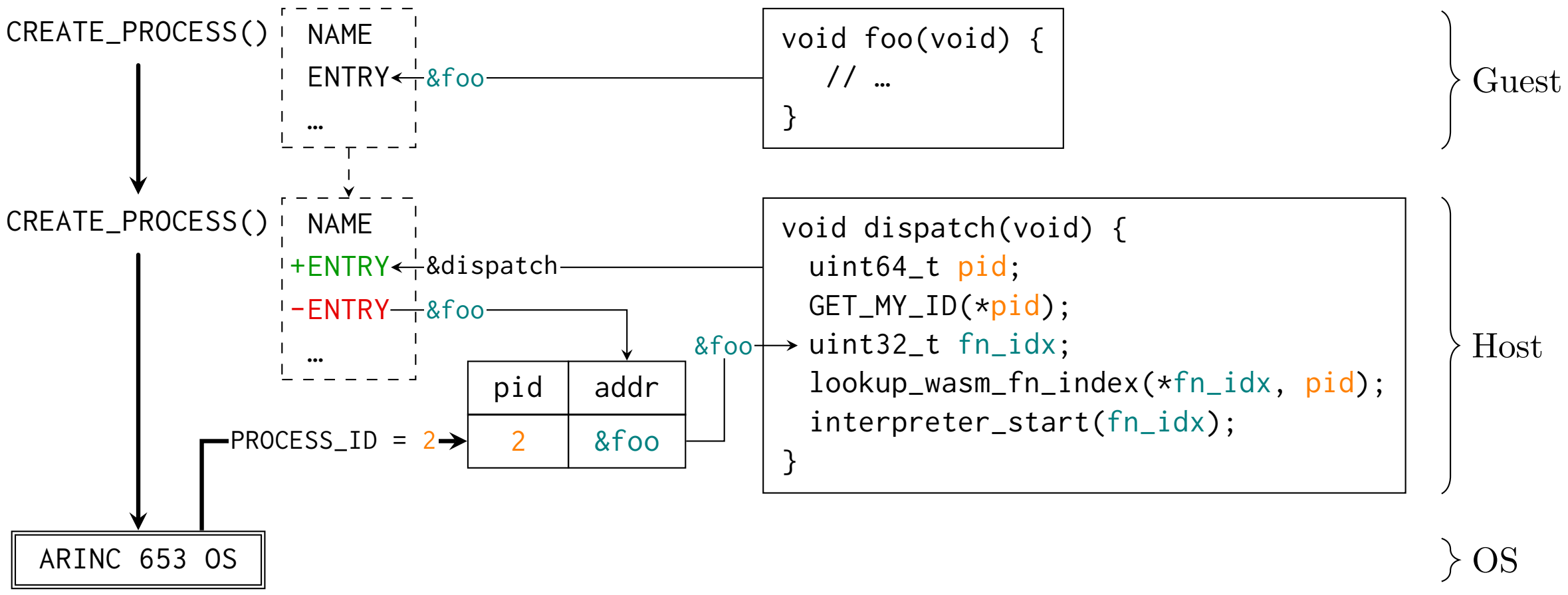

\begin{enumerate}
    \item Guest environment calls \lstinline|CREATE_PROCESS|\textsubscript{Host}, passing a function pointer to the new process's entry point in the \lstinline|PROCESS_ATTRIBUTE_TYPE|\textsubscript{Guest} struct, thus control flows to host environment.
    \item Host environment calls \lstinline|CREATE_PROCESS|\textsubscript{\ac{OS}} itself; however, passing the singular dispatcher function as entry point.
        Control flows to the \ac{OS} and back.
        The new process's process id is returned to the host environment in case of success.
    \item In a static (thus shared across all processes within the current partition) lookup map, using the process id from step 2 as key, the \ac{Wasm} function index (passed as entry function pointer in step 1) is stored as value.\\
    \textbf{Note}:
        \lstinline|CREATE_PROCESS|\textsubscript{\ac{OS}} can only be successfully called during the \lstinline|COLD_START| and \lstinline|WARM_START| modes, during which no process other than the init process is eligible for scheduling.
        This implicates synchronization, eliminating data races to the lookup map --- only a single process can be active during these modes.
    \item When the \ac{OS} actually starts a process (e.g. after a call to \lstinline|START|), control flows into the host environment's singular dispatcher function.
        It in turn calls \lstinline|GET_MY_INDEX|\textsubscript{\ac{OS}} to get it's own process id, and then uses that id to lookup the corresponding function index for the entry function within the guest code.
    \item Host environment invokes that guest code's entry function in a \ac{VM}, flowing control to the guest environment.
\end{enumerate}

The \lstinline|CREATE_ERROR_HANDLER| function can be handled equally, but the need for a lookup map is alleviated; there can only ever be one active error handler function\footnote{Additionally, the error handler can only be set once, and also only during partition initialization, so again no memory synchronization is required \cite[ch. 3.8.2.2]{ARINC653P1_5}.}.

\subsection{Provisioning of Per-Process Storage/TLS}
\label{subsec:init_per_process_storage}

ARINC~653 prescribes shared memory between processes of a single partition~[\nameref{req:HL-3}], necessitating the use of threads in \ac{Wasm} (refer to \autoref{tab:arinc653_partition_vs_process} for a comparison of the terms \textit{thread} and \textit{process}).
When spawning new \ac{Wasm} threads, thread specific setup is required, in particular the linear stack and \ac{TLS}~\cite{wasm_c_abi}.
Recalling that the threading implementation of \ac{Wasm} relies on having quasi-separate \ac{VM} instances that only share the same linear memory (see \autoref{subsubsection:threading}), a problem emerges: without intervention, all threads use the same chunk of the linear memory for their linear stack.
Normally, these details are addressed by the underlying \texttt{pthreads} implementation found for example in a toolchain for the \textit{wasm32-unknown-wasi} target.
However, ARINC~653 provides its own mechanism for processes/threads, and thus we cannot utilize such an external library without compromising the goal of avoiding further dependencies laid out in \autoref{sec:requirements}.

To solve the issue, we provide two functions, to allocate and free per-thread resources.
\autoref{lst:apex_wasm_proc} shows the underlying data structure: a fixed-size array of slots, each holding a thread's linear stack and thread-local storage, alongside an atomic usage marker per slot.
\lstinline|__apex_wasm_proc_alloc| atomically claims the first free slot and points the \lstinline|__stack_pointer| and \lstinline|__tls_base| globals into it; \lstinline|__apex_wasm_proc_free| invalidates both pointers and releases the slot.

\begin{lstlisting}[float, style=C, label=lst:apex_wasm_proc, caption={Per-process linear stack and \ac{TLS} slot allocation. The atomic slot search and global assignment (\lstinline|__stack_pointer|, \lstinline|__tls_base|) are implemented in inline assembly, omitted here for brevity.}]
// per-thread linear stack and TLS, placed in linear memory
__asm__(".globaltype __apex_wasm_proc_ptr, i32\n"
        "__apex_wasm_proc_ptr:\n");
struct __apex_wasm_proc {
  unsigned char tls[APEX_WASM_TLS_SIZE]; // thread-local storage
  // linear stack (after TLS, so a stack overflow corrupts this thread's own
  // TLS first)
  unsigned char ls[APEX_WASM_LS_SIZE];
};

// array of per-thread global state holding structs
struct __apex_wasm_proc __apex_wasm_proc_slots[SYSTEM_LIMIT_NUMBER_OF_PROCESSES];

// True if this slot is currently used
_Atomic _Bool __apex_wasm_proc_usage_markers[SYSTEM_LIMIT_NUMBER_OF_PROCESSES];

// atomically claims a free slot, sets __stack_pointer / __tls_base
__attribute__((export_name("__apex_wasm_proc_alloc")))
_Bool __apex_wasm_proc_alloc(void) { /* inline assembly */ }

// invalidates __stack_pointer / __tls_base, releases the slot
__attribute__((export_name("__apex_wasm_proc_free")))
void __apex_wasm_proc_free(void) { /* inline assembly */ }
\end{lstlisting}

These functions' implementation bears a bootstrapping issue due to a cyclic dependency.
At least when compiling in debug mode/without optimization, LLVM emits \ac{Wasm} instructions that access the linear stack.
Naturally, these instruction must never be executed before the current thread got a distinct linear memory allocated to itself.
Implementing the very function that allocates the linear stack in C would thus access the linear memory before it was allocated and initialized.

There is one escape hatch: for a function whose entire body comprises only inline assembly, even in debug mode without optimization LLVM emits no instructions accessing the linear stack.
Therefore, we opted to implement both functions using only inline assembly.
This proved complicated due to implementation details of LLVM: inline assembly in C using the \lstinline|__asm()__| notation does not use the standardized \ac{WAT} format.
Due to constraints on the assembly parser infrastructure of LLVM, instead an undocumented custom representation is used.
To better understand this \textit{\acl{Wasm}-LLVM-Assembly}, one can dump the corresponding assembly for a given C program using clang's \lstinline|-S| flag.
Further on, semi-relevant documentation is present in the GCC compiler's manual chapters about inline assembly, which in great detail describes most aspects of the syntax.
It however lacks any reference to \ac{Wasm} itself, as the current GCC 15.2 does not support compiling to \ac{Wasm} altogether.

Our implementation remains incomplete, in that it does not currently support allocating differently sized linear stacks for individual threads.
This simplification is in contradiction to the ARINC~653 \ac{API} which explicitly lets the creator of a process pick that processes stack size.
However, considering that during \ac{Wasm} execution the linear stack is second to the \acs{VM}'s internal operand stack, further detailing of stack usage aspects is necessary anyhow.
Lastly, we were surprised to see that LLVM optimizes even our hand-written inline assembly on high optimization level, transparently changing the representation of data in internal variables\footnote{To be precise, this originates from LLVM automatically calling binaryen's \texttt{wasm-opt} on \ac{Wasm} modules unless opted out via the \texttt{--no-wasm-opt} flag.}.

\subsection{Porting the ARINC~653 \texorpdfstring{\ac{APEX}}{APEX} Interfaces to Wasmtime and \texorpdfstring{\ac{WAMR}}{WAMR}}
\label{sec:porting-the-arinc653-interfaces}


This section outlines how to provide a \ac{Wasm} code execution environment as required by \nameref{req:HL-1}.
For a \ac{Wasm} \ac{VM} to be suitable, it must also be executable in an ARINC~653 compliant \ac{OS} in order to meet the requirement \nameref{req:HL-2}.
Furthermore, support for the \ac{Wasm} threads proposal is necessary in order to provide a \ac{TLS}, as discussed in \autoref{subsec:init_per_process_storage}.

The \acl{Wasm} Working Group requires two independent implementations of a new \ac{Wasm} proposal\footnote{\url{https://www.w3.org/2023/wasm-wg-charter.html}, see `Success Criteria'}, so two representative (embedded) \ac{Wasm} \ac{VM}s --- namely Wasmtime and \ac{WAMR} --- were chosen.
\ac{WAMR} was\cite{wamr_moves_away_from_bytecodealliance}, and Wasmtime is officially maintained by the Bytecode Alliance.
However, to ease the implementation, as much code as possible is written \ac{Wasm}-\ac{VM}-agnostic to enable re-use (even with other \ac{Wasm} \acp{VM}).

The \ac{Wasm} specification defines so-called \textit{host functions}\cite[sec.~4.2.6]{WebAssemblyCoreSpecification2}, coupling function imports and exports of a \ac{Wasm} module to the host environment.
Consequently, the \ac{VM} must provide these imports and may use the exports.
While some \ac{Wasm} \acp{VM} feature plugin mechanisms to load additional modules at startup (such as WasmEdge), neither Wasmtime nor \ac{WAMR} support this.
Thus, dedicated \acp{CLI} were established based on Wasmtime's C \ac{API} and \ac{WAMR}'s static \texttt{libiwasm}.
Notably, the two \acp{VM} differ in how shared linear memory is configured: \ac{WAMR} enables it by default, whereas Wasmtime requires explicit setup via its shared-memory \ac{API}~\cite[sharedmemory.h]{wasmtime_c_api}.

To support any ARINC~653 library without modification, the registered host functions are solely wrappers mediating between the \ac{Wasm}-guest-exposed ARINC~653 \ac{APEX} interfaces and a host-environment-provided ARINC~653 library.
Struct field access across the guest/host \ac{ABI} boundary (including endianness and 32-bit/64-bit width differences) is handled by \texttt{c-abi-lens} as described in \autoref{sec:abi_translation}.
Function pointer handling follows the dispatch mechanism described in \autoref{sec:handling_of_function_pointers}; notably, the index-based representation suits ARINC~653 \ac{APEX} code, but could be problematic under the \ac{Wasm} Component Model where function tables may change at runtime.

Unfortunately, there is no canonical representation for function signatures of \ac{Wasm} \ac{VM} host functions, so either \ac{VM} defines its own.
The wrappers require all function parameters supplied by the \ac{Wasm} guest, as well as the base address of \ac{Wasm}'s linear memory (illustrated for Wasmtime in \autoref{lst:host_func_wrapper_wasmtime} and \ac{WAMR} in \autoref{lst:host_func_wrapper_wamr} with error handling omitted for brevity).
In Wasmtime, the host function follows \lstinline|wasmtime_func_unchecked_callback_t|~\cite{wasmtime_c_api}; in \ac{WAMR}, \lstinline|wasm_runtime_register_natives_raw|~\cite{wamr_export_file_reference_archived} provides one wrapper parameter holding all host function parameters as a list, from which macros retrieve specific arguments.

\begin{lstlisting}[float, style=C,label=lst:host_func_wrapper_wasmtime, caption={Wasmtime host function definition.}, deletekeywords={extern}]
wasm_trap_t *<wasmtime_host_function_wrapper>(
  void               *env,
  wasmtime_caller_t  *caller,
  wasmtime_val_raw_t *args_and_results,
  size_t              num_args_and_results
) {
  /* 1. Get linear memory export from caller */
  wasmtime_extern_t ext;
  wasmtime_caller_export_get(caller, "memory", strlen("memory"), &ext);

  /* 2. Get base address of linear memory */
  wasmtime_context_t *context =
    wasmtime_caller_context(caller);

  uint8_t *wasm_linear_memory_baseaddr =
    wasmtime_memory_data(context, &ext.of.memory);

  /* 3. Decode arguments, perform host-side work, write results ... */
  ...

  /* NULL trap => success */
  return NULL;
}
\end{lstlisting}

\begin{lstlisting}[float, style=C, label=lst:host_func_wrapper_wamr, caption={\ac{WAMR} host function definition.}]
void <wamr_host_function_wrapper>(
  wasm_exec_env_t exec_env,
  uint64_t       *args
) {
  /* 1. Get module instance for this execution environment */
  wasm_module_inst_t module_inst =
    wasm_runtime_get_module_inst(exec_env);

  /* 2. Get base address of linear memory (app space -> native) */
  uint8_t *wasm_linear_memory_baseaddr =
    wasm_runtime_addr_app_to_native(module_inst, 0);

  /* 3. Decode arguments, perform host-side work, write results ... */
  ...
}
\end{lstlisting}

To inform each \ac{Wasm} \ac{VM} of the number and type of parameters per host function, the implementation uses \ac{WAMR}'s function-signature convention for both \acp{VM}.
A C-string is employed in which each letter within \lstinline|()| describes a parameter and the letter after \lstinline|)| the return type~\cite{wamr_export_native_api.md}.
For example, \lstinline|(iIiiI)| represents five parameters and no return value, where lowercase \lstinline|i| denotes a 32-bit integer and uppercase \lstinline|I| a 64-bit integer.
For Wasmtime, this signature string is converted at startup into \lstinline|WASM_I32| and \lstinline|WASM_I64| parameter representations, stored in a \lstinline|wasm_valtype_vec_t|.

\begin{lstlisting}[float, style=C,
  label=lst:wamr_native_symbol,
  caption={\ac{WAMR} definition of a native symbol.}]
typedef struct NativeSymbol {
  const char *symbol;
  void       *func_ptr;
  const char *signature;
  void       *attachment;
} NativeSymbol;
\end{lstlisting}

For each ARINC~653 \ac{APEX} function, a wrapper host function calls the corresponding \ac{APEX} function within the host-environment-provided ARINC~653 library.
\ac{WAMR}'s \lstinline|struct NativeSymbol| (\autoref{lst:wamr_native_symbol}) bundles the function name, address, and signature~\cite{wamr_export_native_api.md};
applying the same array structure to Wasmtime --- with a macro-driven conversion of the signature string --- facilitates a single shared registration array to serve both \acp{VM}.

\section{Evaluation}
\label{sec:evaluation}

The aforementioned culminated in two distinct repositories demonstrating various aspects to assert the compliance with the requirements from \autoref{sec:requirements}, DLR-FT/arinc653-wasm\footnote{\url{https://github.com/DLR-FT/arinc653-wasm/}} and airbus/a653lib\footnote{\url{https://github.com/airbus/a653lib}}.
Requirements from \autoref{sec:requirements} are verified as following:

%

\begin{itemize}
    \item \nameref{req:SYS-1} compliance is demonstrated twice independently: the repository of \textit{AIRBUS' a653lib} contains a \texttt{CMakeLists.txt}, and the \textit{DLR-FT arinc653-wasm} repository a \texttt{Makefile} both suitable to compile partitions to \ac{Wasm}.
    \item \nameref{req:SYS-2} compliance is demonstrated in the AIRBUS a653lib repository, using Linux combined with a653lib itself as the ARINC~653 \ac{OS}.
    \item \nameref{req:SYS-3} is not demonstrated for lack of a second ARINC~653 \ac{OS}.
    \item \nameref{req:HL-1} is demonstrated in AIRBUS' a653lib.
        Additionally DLR-FT arinc653-wasm's \texttt{arinc653-wasm-harness} provides a minimal execution environment offering a subset of ARINC 653 functionality\footnote{The harness only implements a bare subset of ARINC~653, foremost to verify process creation}.
    \item \nameref{req:HL-2} is demonstrated twice via the \texttt{p\_wamr} and \texttt{p\_wasmtime} partitions part of AIRBUS' a653lib repository.
    \item \nameref{req:HL-3} is not fully demonstrated; while the \ac{API} offered in all three environments (\texttt{p\_wamr}, \texttt{p\_wasmtime} and \texttt{arinc653-wasm-harness}) strives to adhere to ARINC~653, it is neither complete; nor are the authors confident that it is completely compliant.
        The entire ARINC~653 \ac{API} surface is large and complex; while we cover all corner cases known to us (such as passing of function pointers), assurance of full coverage of the \ac{API} escapes this paper's scope.
    \item \nameref{req:HL-4} is demonstrated in the form of the AWK script found in DLR-FT arinc653-wasm repository, based upon the rationale statements found in \autoref{sec:arinc653_wasm_headers} and \autoref{apx:rationale}, which document all decisions regarding implementation dependent aspects of the header files.
    \item \nameref{req:HL-5} is not demonstrated for the same reason as \nameref{req:SYS-3}.
    However, our analysis of the resulting \ac{Wasm} files indicates independence from the \ac{OS} due to the lack of non ARINC~653 service imports\footnote{
    Some of the \ac{Wasm} files do import \ac{WASI} functions due to using libc \acp{API} (e.g. \lstinline|snprintf|) beyond what ARINC~653 offers, disciplined use of libc imports reduced to the freestanding environment\cite[ch. 4]{ISO9899} would resolve this.}.
\end{itemize}

Further on, the abstract promises three main contributions, which we ought to revisit here:

\begin{enumerate}
    \item An ambiguity-free ARINC~653 header file is provided via the \texttt{include/ARINC653-wasm.h} target in DLR-FT arinc653-wasm's \texttt{Makefile}.
        Rationale statements for all decisions leading to the specific header file are documented in \autoref{apx:rationale} and further explained in \autoref{sec:arinc653_wasm_headers}.
    \item A \ac{Wasm} based \ac{ABI} is provided and exercised twice in the DLR-FT and the AIRBUS repositories.
    \item Using AIRBUS' a653lib, both Wasmtime and \ac{WAMR} can be used to execute ARINC~653 partitions compiled to \ac{Wasm}.
        This executes both on the x86\_64 and the aarch64 \acp{ISA}, using the exact same \texttt{partition\_\{a,b\}.wasm} binaries.
\end{enumerate}


As noteworthy contribution on the side, we provide the C-\acs{ABI}-lens, a code generator enabling retrieval and mutation of structs available in memory which however are represented in a foreign \ac{ABI}'s memory layout.

Finally, for testing purposes the \texttt{arinc653-wasm-harness} provides a simple environment providing a subset of ARINC~653's \ac{API}, while transparently mapping sampling ports to IP/UDP traffic.
It is implemented fully agnostic of the \ac{OS}, and thus works on Windows, Linux and Mac OS alike.

\section{Limitations and Future Work}
While our proof-of-concept implementation satisfies the promise from the abstract, several open points remain as follows.

\subsection{Representation of Messages from Physical \texorpdfstring{\acs{I/O}}{I/O} Interfaces}
Most ARINC~653 \acp{OS} use sampling and queuing ports to connect partitions to physical \ac{I/O} interfaces (such as ARINC~664 or ARINC~825).
For example, the specific representation of ARINC~825 frames received via a sampling port remains \ac{OS} vendor dependent.
This results in a tight coupling between each ARINC~653 partition's interfacing with physical \ac{I/O} and specific \ac{OS}'s drivers.
Such coupling remains as a challenge when \ac{Wasm} is used as a partition compilation target.
An effective mitigation is the use of dedicated \ac{I/O} partitions for pre-/post-processing of messages, however at the cost of additional latency and computational overhead.

\subsection{Provide Ada support}
Activities were carried out to provide Ada language stubs employing the ARINC~653 C-headers.
Furthermore, the GNAT-LLVM toolchain has been set up to compile ARINC~653 enabled Ada source code to \ac{Wasm}.
Further investigations are needed to determine whether pure language stubs are sufficient to enable Ada.
These activities have not been upstreamed yet.
        
\subsection{Measure Performance Impact}
Virtualization (and especially interpretation) is not free, and often incurs a performance penalty both in execution time and memory footprint.
Any serious use of this paper's concept requires elaborate measurements to assess the extent of these overheads.
Even in the case of \ac{AOT} compilation, loads and stores are bound-checked to stay within \ac{Wasm}'s linear memory, thus a performance penalty is likely.

\subsection{Port to an Execution Environment Suitable for Safety-critical Systems}
The current demonstration runs on Linux, which, despite growing efforts (e.g. the ELISA aerospace working group\footnote{\url{https://github.com/elisa-tech/wg-aerospace}}), is not yet an accepted \ac{OS} for safety-critical avionics.
Ideally, the demonstrator is ported to one of the existing industrial ARINC~653 compatible \acp{OS} mentioned in \autoref{sec:introduction}.
At the bare minimum, a port to a bare-metal environment without the modern amenities of a full \ac{OS}, such as a file-system and syscalls to (re-)map memory pages, would strengthen the demonstration's impact.

\subsection{Utilization of a \texorpdfstring{\acs{Wasm}}{Wasm} Interpreter Tailored for use in Safety-critical Systems}
As is, the demonstration employs Wasmtime and \ac{WAMR}, two popular \ac{Wasm} \acp{VM} backed by the Bytecode Alliance and/or large communities.
However, neither is made with safety standards such as ED-12C/DO-178C in mind, and particular design choices --- such as the generation of object code at runtime for increased performance, or rewriting to a custom intermediate representation optimized for efficient interpretation --- may hinder eventual certification activities~\cite{zaeske_wasm_avionics_2025}.
Especially looking towards high assurance, the use of a simpler, more certifiable \ac{Wasm} interpreter could increase the effort's credibility.
Work on our own interpreter tailored for use in safety-critical avionics\footnote{\url{https://github.com/DLR-FT/wasm-interpreter}} continues.

\subsection{\texorpdfstring{\acs{Wasm}}{Wasm}64 Support}
The current demonstration uses \ac{Wasm}32, the default, which models addresses to the linear memory with 32-bit integers.
However, with the release of the \ac{Wasm} specification 3.0~\cite{wasm3_release_news}, 64-bit addressing has been ratified too.
Implementing this requires no changes to the ARINC~653 header file presented to partition developers, but adaptations to both the build system and the wrapper functions exposed from the host to the guest environment.

\subsection{Eliminating \texorpdfstring{\acs{WASI}}{WASI} altogether}
Currently, the demonstrator uses the \textit{wasm32-unknown-wasi} toolchain, as that comes with a libc implementation which is used for includes such as \lstinline|stdint.h|.
The bare \textit{wasm32-unknown-none} toolchain does not come with the header files of a C standard library, severely limiting usability.
While non-\ac{OS}-specific features like guaranteed bit-width integer types from \lstinline|stdint.h| or string formatting via \lstinline|snprintf| are of great value, \ac{OS}-specific functionality using \ac{WASI} functions should not be used by applications targeting ARINC~653, as it escapes the defined scope of the \ac{OS} \ac{API}.
Currently, wasi-libc is the only libc suitable for use in \ac{Wasm} with sufficient coverage of ISO-C\cite{ISO9899}; a non-\ac{WASI} libc conforming to the freestanding profile \cite[chapter 4]{ISO9899} would be a welcome contribution.

\subsection{Exploring \texorpdfstring{\acs{WIT}}{WIT} to Describe an Interface to ARINC~653}
For the context of this paper, we refrained from using \ac{WIT} altogether.
A detailed analysis of the implications of using \ac{WIT} is, however, warranted, and we suspect the existence of use-cases where it offers substantial advantages over the constellation of C-\ac{ABI}-based workarounds described in this paper.
As this would entail syntactic changes to the ARINC~653 \ac{API} described in ARINC~653 part 1\cite{ARINC653P1_5}, the biggest downside would likely be a loss of source-code compatibility compared to the work laid out in this publication.

\subsection{Convincing the SAE to license the ARINC~653 header files}
The SAE (as the umbrella organization for ARINC) offers ARINC~653 Part 1 and Part 2 C header files for download\footref{footnote:arinc_headers}.
They form the basis for our contribution in \autoref{sec:arinc653_wasm_headers}, but we cannot redistribute the result of that work, as the input files (the header files offered for download) are not provided under any license.

\subsection{Investigate \texorpdfstring{\acs{Wasm}}{Wasm} Fuel-based Temporal Partitioning}

As time-sliced partition execution is a central concern of ARINC~653, the authors further suggest investigating the non-standard \ac{Wasm} Fuel mechanism (available in Wasmtime):
Fuel allows a \ac{Wasm} module to execute until a defined instruction budget is exhausted, after which the module yields\cite{10311207}.
Consequently, an instruction-based \ac{Wasm} Fuel scheduling could replace traditional time-based ARINC~653 partition scheduling, enabling significantly finer granularity for scheduling ARINC~653 partitions.
Early research indicates potential to address multi-core interference issues too\cite{dynamic_mitigation_of_multicore_interference}.

\section{Conclusion}
\label{sec:conclusion}

The memory layout of data types and integer bit widths in ARINC~653 are \ac{OS}, compiler- and \acs{ISA}-dependent, allowing for source-level portability but not object-code compatibility.
\ac{Wasm}, in contrast, enables object-code compatibility independent of any specific hardware architecture.
However, solely providing \ac{APEX}-like host functions as part of a \ac{Wasm} \ac{VM} is insufficient to enable true forward-compatibility.

To establish a stable \ac{ABI} for ARINC~653 partitions, this paper contributes concrete, opinionated \ac{APEX} headers, 32-/64-bit and endianness-aware guest/host handling, and a function pointer guest-to-host pass-through and calling mechanism.
As today's \ac{Wasm} threading remains unstable, a self-managed per-process linear stack and \ac{TLS} slot allocator are additionally proposed, ensuring that \ac{Wasm} ARINC~653 partitions execute as native ARINC~653 processes with the intended shared memory view.
For demonstration purposes, \ac{CLI} implementations were realized atop a Linux-based ARINC~653 scheduler, employing either \ac{WAMR} or Wasmtime as \ac{Wasm} \acp{VM}.

As the ARINC~653 \ac{APEX} definition does not conform to the \ac{WIT} specification (no function pointer pass-through from guest to host; no values returned through argument pointers), the implementation cannot be upstreamed to \ac{WASI}.
Nevertheless, any ARINC~653-enabled partition can be ported to this environment by changing the compilation target alone, as source-level \ac{API} compatibility is fully preserved.
For independent reproduction and experimentation, the implementation is openly available\footnote{\url{https://github.com/airbus/a653lib}}.

\section{Appendix: Design Decisions \& Rationale}
\label{apx:rationale}

\begin{itemize}
\item
  \textbf{Choice}:
  \texttt{\textless{}an\ APEX\ integer\ type\textgreater{}}
  -\textgreater{} \texttt{APEX\_LONG\_INTEGER}.\\
  \textbf{Reason}: These placeholders are only used for
  \texttt{*\_ID\_TYPE}. All of these IDs are only ever handed out by the
  OS. By picking the largest possible type available (64-bit integer),
  we ensure that we are compatible with any possible OS, as we can store
  any possible ID. For an OS that chooses a 32-bit integer, the upper
  bits just remain untouched.
\item
  \textbf{Choice}: import the linear memory (via
  \texttt{-\/-import-memory} linker flag).\\
  \textbf{Reason}: Each partition has multiple processes, which are
  guaranteed to have a shared address space (ARINC 653 P1-5 chapter
  2.3.2). The only way to achieve this is if they have shared linear
  memory. To cause that, they all need to import the linear memory.
\item
  \textbf{Choice}: export the function table (via
  \texttt{-\/-export-table} linker flag).\\
  \textbf{Reason}: In order for the \texttt{CREATE\_PROCESS} call to
  succeed, the host environment needs to be able to call a guest
  environment function identified via an index into said table.
  Exporting the table ensures that the funcref table is accessible from
  the host environment.
\item
  \textbf{Choice}: Do not use \texttt{\_\_funcref\_t} for function
  pointers, e.g.~the \texttt{ENTRY\_POINT} as argument in the
  \texttt{CREATE\_ERROR\_HANDLER} function.\\
  \textbf{Reason}: \texttt{\_\_funcref\_t} is not representable in
  Linear Memory. Hence, it can not become the field of a struct passed
  by-reference. However, the \texttt{PROCESS\_ATTRIBUTE\_TYPE} struct
  comprises an \texttt{ENTRY\_POINT} field holding a function pointer.
  As \texttt{\_\_funcref\_t} can not be used there, it is necessary to
  expose a Wasm table for function pointers. Therefore, any use of
  \texttt{\_\_funcref\_t} shall be avoided, in order to keep all
  function pointer representations consistent. (In theory,
  \texttt{CREATE\_ERROR\_HANDLER} could employ \texttt{\_\_funcref\_t},
  as the function pointer is passed directly as function argument,
  however, a unified handling of function pointers is preferred.)
\item
  \textbf{Choice}: Do not use \texttt{\_\_externref\_t} for external
  resources, e.g.~\texttt{SAMPLING\_PORT\_ID} as argument to the
  \texttt{CREATE\_SAMPLING\_PORT} function.\\
  \textbf{Reason}: \texttt{\_\_externref\_t} is used to pass references
  to host environment objects through the guest environment, keeping
  them opaque to the guest environment. It is however desirable to have
  these accessible from the guest environment, i.e.~for a partition to
  report back a sampling port's ID. Further, making opaque what APEX
  discloses, would be a breaking change from the APEX API.
\item
  \textbf{Choice}: Use \texttt{\_\_wasm\_\_} preprocessor
  \texttt{\#define} to conditionally annotate official APEX header
  functions with \texttt{import\_module} \& \texttt{import\_name}
  attributes.\\
  \textbf{Reason}: Clang is the sole compiler, that is capable to
  compile C to Wasm reasonably. Heaving two headers, one without and one
  with Wasm specific annotations, is unfavorable.
\item
  \textbf{Choice}: Use \texttt{RETURN\_CODE\_TYPE} -\textgreater{}
  \texttt{i32}.\\
  \textbf{Reason}:
  \href{https://www.open-std.org/jtc1/sc22/wg14/www/docs/n2310.pdf\#subsubsection.6.7.2.2}{Chapter
  6.7.2.2 of ISO/IEC 9899:2018 (``C17'')} states that the representation
  of enum types is implementation dependent. However, there is not much
  of a choice here, since the
  \href{https://github.com/WebAssembly/tool-conventions/blob/main/BasicCABI.md}{Wasm
  C ABI} dictates \texttt{enum}s to be represented as \texttt{i32}.
  \href{https://www.open-std.org/jtc1/sc22/wg14/www/docs/n3220.pdf\#subsubsection.6.7.3.3}{Chapter
  6.7.3.3 of ISO/IEC 9899:2024 (``C24'')} introduces the possibility of
  specifying the underlying type for \texttt{enum}s, which we
  intentionally ignore to maintain compatibility with older versions of
  the C standard.
\item
  \textbf{Choice}: Employ manual
  \href{https://clang.llvm.org/docs/AttributeReference.html\#import-module}{\texttt{import\_module}}
  \&
  \href{https://clang.llvm.org/docs/AttributeReference.html\#import-name}{\texttt{import\_name}}
  attributes in C-header file instead of using
  \href{https://component-model.bytecodealliance.org/design/wit.html}{WIT}.\\
  \textbf{Reason}: Although WIT is the preferred approach to generate
  language bindings, \texttt{wit-bindgen} neither guarantees (nor does
  it) generate APEX compliant C header files (structs, typedefs and
  foremost function signatures). As APEX is merely used by C, C++ and
  ADA (all of which falling back to the C ABI as FFI), WIT is neither
  applicable nor required.
\item
  \textbf{Choice}: Use \texttt{\_\_wasm\_\_} preprocessor macro to
  decide insertion of \texttt{import\_module} \& \texttt{import\_name}
  attributes into the APEX header.\\
  \textbf{Reason}: This allows the header file to be used for non-Wasm
  compilation too, opening the door for future experiments. At the same
  time, compatibility with other emerging compilers is increased, as
  they may use different syntax for Wasm import annotations.
\item
  \textbf{Choice}: Return values through pointer function arguments.\\
  \textbf{Reason}: WebAssembly is influenced by modern-ish languages
  like OCaml \& Rust, which in part favor returning even complex data
  types (like structs) from functions by-value. The APEX interfaces
  however rely on pass-by-reference (pointer to input and output values)
  for the majority of arguments. This is aligned with
  \href{https://github.com/WebAssembly/tool-conventions/blob/main/BasicCABI.md\#function-arguments-and-return-values}{WebAssembly
  Tool Conventions' Basic C ABI} which for structs and unions larger
  than one element also prescribes pass by-reference (denoted
  \emph{indirect}).
\end{itemize}

\printbibliography

\section*{AI Use Disclosure}
The authors used AI-assisted tools (DeepL Write, Claude Sonnet 4.6) for language editing and proofreading of this manuscript.
All scientific content, technical analysis, and conclusions are solely the responsibility of the authors.

\end{document}